\documentclass[useAMS,usenatbib,usegraphicx,12pt]{mn2e}
\usepackage{array,multirow,epsfig,graphicx,amssymb,amsmath,lscape}
\usepackage{hyperref,txfonts,rotating,psfrag}
\usepackage{aas_macros}

 \title{Variability of black hole accretion
discs:  The cool, thermal disc component}

\author[M. Mayer \& J. E. Pringle]{M. Mayer$^1$ \& J. E. Pringle\\
Institute of Astronomy, Madingley Road,
Cambridge CB30HA, UK\\
$^1$ E-Mail: mm@ast.cam.ac.uk}

\begin{document}
\newcommand{\cm}{\textrm{ cm}}
\newcommand{\g}{\textrm{ g}}
\newcommand{\K}{\textrm{ K}}
\newcommand{\jw}{\textrm{j/w}}
\date{\today }

\pagerange{\pageref{firstpage}--\pageref{lastpage}} \pubyear{2005}

\maketitle
\label{firstpage}

\begin{abstract}
We extend the model of \citet{2004MNRAS.348..111K} for variability in
black hole accretion discs, by taking proper account of the thermal
properties of the disc. Because the degree of variability in the
\citet{2004MNRAS.348..111K} model depends sensitively on the ratio of
disc thickness to radius, $H/R$, it is important to follow the
time-dependence of the local disc structure as the variability
proceeds. In common with previous authors, we develop a one-zone model
for the local disc structure. We agree that radial heat advection
plays an important role in determining the inner disc structure, and
also find limit-cycle behaviour. When the stochastic magnetic dynamo
model of \citet{2004MNRAS.348..111K} is added to these models, we find
similar variability behaviour to before. 

We are now better placed to put physical constraints on
model parameters. In particular, we find that in order to be
consistent with the low degree of variability seen in the thermal disc
component of black hole binaries, we need to limit the energy density of the
poloidal field that can be produced by local dynamo cells in the disc
to less than a few percent of the energy density of the dynamo field
within the disc itself.

\end{abstract}


\section{Introduction}

Accretion powered x-ray sources, both on the galactic (AGN) and
stellar (X-ray binaries) scales, display significant aperiodic
variability, or flickering, on a broad range of timescales \citep[see
for
example][]{1988MmSAI..59..239M,1990A&A...230..103B,1994ApJS...92..511V,
2003ApJ...593...96M,2003MNRAS.345.1271V,0306213}. The origin of this
variability is not well understood. However,
\citet{2004MNRAS.347L..61U} and \citet{2005MNRAS.359..345U} have
pointed out that the existence of a strong linear relationship between
the amplitude of the X-ray variability and the amplitude of the x-ray
flux \citep{2001MNRAS.323L..26U} -- the rms-flux relation -- can be
used as a diagnostic to distinguish between various models. For
example \citet{2005MNRAS.359..345U} emphasise that the standard simple
shot-noise models, where the light curve is produced by the summation
of randomly occurring shots, or flares, cannot explain such a relation.
Both \citet{2004MNRAS.347L..61U} and \citet{2005MNRAS.359..345U} note
that such a relation can be produced naturally by the model of
\citet{1997MNRAS.292..679L} where the variability is produced by
variations in the accretion rate occurring at different radii which
propagate inwards, and so modulate the energy release in the central
x-ray emitting region. The major problem with this model, as noted by
\citet{1997MNRAS.292..679L}, is the physical origin of the accretion
rate variations. In order for the model to work, as also emphasised by
\citet{2001MNRAS.321..759C}, it is necessary for the timescale of the
variations at each radius to be at least as long as the viscous
timescale at that radius. The most obvious origin for fluctuations in
an accretion disc are turbulent or hydro-magnetic, perhaps associated
with the disc dynamo, and these take place on the local dynamical
timescale ($\sim \Omega^{-1}$, where $\Omega$ is the angular
velocity), which is shorter than the local viscous timescale by a
factor of $\sim \alpha (H/R)^2$, where $\alpha$ ($<$1) is the
viscosity parameter, and $H/R < 1$ the disc opening angle
\citep{1981ARA+A..19..137P}.

A solution to this problem, in the form of an explicit physical model,
has been recently proposed by \citet{2004MNRAS.348..111K}, based on
ideas put forward by \citet{2003ApJ...593..184L}. They suggest that
the local dynamo processes in the disc can affect the accretion rate
by driving angular momentum loss in the form of an outflow (wind or
jet). They model the dynamo as a small-scale stochastic phenomenon,
operating on roughly the local dynamical timescale. They then postulate
that large-scale outflow can only occur when small-scale random
processes in neighbouring disc annuli give rise by chance to a
coherent large-scale field. This occurs on much longer timescales, by
a factor of order $2^{R/H}$ \citep{2003ApJ...593..184L}. This also
provides an explanation for why the dominant flickering frequencies
are typically much longer than the dynamical timescales at the centre
of the disc where most of the energy is released.

However, \citet{2004MNRAS.348..111K} took a very idealised model for
the disc in that they assumed that the disc had a constant thickness
ratio of $H/R$. This simplified the computations for two main
reasons. First, local disc structure equations could be ignored,
because there was no need to compute the local value of disc
thickness. Second, because the local disc dynamos were assumed to be
spatially independent on a radial scale of $\Delta R \sim H$, there
was no need to vary the number of independent dynamos as the disc
thickness varied with time. In this paper we take the first steps
towards remedying this deficiency. For the time being we concern
ourselves with a standard (optically thick, geometrically thin)
accretion disc, and thus any conclusions we have will be relevant
mainly to
the thermally dominated (TD) state of the X-ray transients
\citep[e.g.][]{0306213}. We leave the extension to the more
interesting, and more variable (low/hard) state, containing both disc
and corona, to future work. In Section~\ref{sect:methods} we explain
how we generalise the work of \citet{2004MNRAS.348..111K} by including
equations to compute the local disc structure, and also explain how we
vary the grid spacings to follow the radial size of the local disc
dynamo cells in such a way as to minimise unwanted numerical
mixing. In Sections~\ref{sect:stationary} and~\ref{sect:timedep} we
investigate the models which result from our equations when there is a
steady external accretion rate, and in Section~\ref{sect:review}
compare our results with previous work in the field. In
Section~\ref{sect:flickering} we add the variability according to the
stochastic magnetic model of \citet{2004MNRAS.348..111K}. In
Section~\ref{sect:obs}, we discuss the observational data on the
variability of the thermal disc component, to which our models are
relevant, and in Section~\ref{sect:res} we show that for expected
values of the model parameters our models are consistent with the
data. We present brief conclusions in Section~\ref{sect:conclusions}.

\section{Methodology} \label{sect:methods}

In this Section we present the input physics for our basic accretion
disc models. For a more detailed discussion, see for example
\citet{2002apa..book.....F}.

 The disc is treated as one-dimensional, in the sense
that we resolve the disc in radial direction only, dividing the disc
into annuli, and just use the one-zone approximation in the vertical
direction.  Thus, all variables (e.g. mass density $\rho$, temperature
$T$ etc.) have one value at each radius $R$, and we make no further
assumptions about the vertical disc structure. Thus, for example the
surface density $\Sigma$ is given in terms of density $\rho$ and disc
semi-thickness $H$ by the relation
\begin{equation}
  \label{eq:sigma}
  \Sigma = 2\rho H
\end{equation}
As we discuss in Section~\ref{sect:review} it is possible to make more
detailed assumptions about the vertical disc structure. However, in
view of the large number of uncertainties in the basic physical
processes involved here, we  regard such extra complications as
unnecessary for our current purposes.

\subsection{Surface density} \label{sect:surfdens}

The viscous evolution of the disc is governed
\citep[][]{1981ARA+A..19..137P,1992MNRAS.259P..23L} by
the equation of continuity

\begin{equation}
\label{eq:kg1}
\frac{\partial 
\Sigma}{\partial t}+\frac{1}{R}\frac{\partial}{\partial R}\left( \Sigma R
  u_R\right) = 0
\end{equation}
and by the equation of conservation of angular momentum
\begin{equation}
\label{eq:angmom1}
  \frac{\partial}{\partial t}\left(\Sigma
    R^2\Omega_\textrm{K}\right)+\frac{1}{R}\frac{\partial}{\partial
    R}\left(\Sigma R u_R R^2\Omega_\textrm{K}\right) = \frac{1}{2\pi
    R}\frac{\partial G}{\partial R}.
\end{equation}
Here $\Sigma$ is the surface density, $u_R$ the radial velocity, $\dot
M=2\pi \Sigma R u_R$ the accretion rate, $\Omega_\textrm{K}=\sqrt{GM/R^3}$ the Keplerian
rotation frequency and $G$ the viscous torque. 

We should note two things here. First, although we are discussing
accretion discs around black holes in this paper, we use purely
Newtonian gravity, around a point mass $M$, and truncate the disc at
an appropriate radius $R_{\rm in} = 6GM/c^2$. Here again we take the
view that the wealth of uncertainties inherent in the physical
processes involved, and the complicated nature of the subsequent
behaviour, means that adding the complication of using the proper
space-time geometry is not warranted at this stage. Second, by taking
the angular velocity of the disc material to be
$\Omega_\textrm{K}=\sqrt{GM/R^3}$ we have also neglected the effects of
any radial pressure gradient in the disc. We shall show below that
this approximation is justified for the disc models presented here.

As is apparent from equation~\ref{eq:angmom1}, angular momentum is
transported either by advection or viscous torques.  For Keplerian
rotation the viscous torque is given by
\begin{equation}
  \label{eq:torque}
  G=-3\pi \nu_\textrm{t} \Sigma R^2\Omega_\textrm{K},
\end{equation}
where $\nu_\textrm{t}$ is the kinematic viscosity.

Equations (\ref{eq:angmom1}) and (\ref{eq:kg1}) can be combined
\citep{1981ARA+A..19..137P} to give

\begin{equation}
\label{eq:angmom}
\frac{\partial \Sigma}{\partial t}=\frac{3}{R}\frac{\partial}{\partial
  R}\left[R^{1/2}\frac{\partial}{\partial R}\left(\nu_\textrm{t} \Sigma
    R^{1/2}\right)\right]. 
\end{equation}

We integrate this equation in time applying a zero-torque condition at
the inner boundary ($R_\textrm{in}=3R_\textrm{S}$, where $R_\textrm{S}
= 2GM/c^2$) and thus set $\Sigma_{R_\textrm{in}=3R_\textrm{S}}=0$. At
the outer boundary we feed the disc with a constant external accretion
rate $\dot M_\textrm{ext}$. For advection we use a first-order donor
cell upwind scheme. The integration conserves mass up to machine
accuracy, accounting for mass supply at $R_\textrm{out}$ and loss at
$R_\textrm{in}$.

\subsection{Radial force balance and hydrostatic equilibrium} 
\label{sect:hydrostat}

We assume throughout that the disc is in hydrostatic equilibrium in
the vertical direction (i.e. perpendicular to the disc plane). In the
radial direction, the central gravitational pull is balanced by the
centrifugal force
\begin{equation}
\label{eq:radforce}
\Omega_\textrm{K}^2 R=\frac{GM}{R^2}.
\end{equation}

The vertical structure is treated in the one-zone-approximation. Thus
we take the vertical component of the gravitational force of the black
hole to be balanced by the vertical pressure gradient, i.e.
\begin{equation}
\frac{\partial P}{\partial z} = - \rho g_z,
\end{equation}
where $g_z$ is the vertical gravity. In terms of our one-zone model we
may write
\begin{equation}
\frac{\partial P}{\partial z} = P/H,
\end{equation}
and
\begin{equation}
g_z = -\Omega_\textrm{K}^2 R \frac{H}{R}.
\end{equation} 
Using this, together with Equation~\ref{eq:sigma} we obtain the
relationship expressing vertical hydrodynamic equilibrium in the form
\begin{equation}
\label{eq:hydrostat}
P=\frac{GM}{4\rho R^3}\Sigma^2. 
\end{equation}

\subsection{Energy equation} \label{sect:temp}

We consider the heat content $q$ of one half of an elemental disc
annulus of width $\Delta R$ at radius $R$, and height $H$. Then, from
the usual thermodynamic relations, and assuming hydrostatic
equilibrium (Equation~\ref{eq:hydrostat}), we find (see
Appendix~\ref{sect:appenergy}) that
\begin{equation}
\label{eq:energy}
\frac{dq}{dt}=\dot e + u_R\frac{\partial e}{\partial R} +A P \dot H+ P \frac{\partial \left(2\pi R H u_R\right)}{\partial
  R}.
\end{equation}
Here $e$ is the internal energy of the semi-annulus, and the last two
terms come from the $'PdV'$ work done on the semi-annulus. $A=2\pi R
\Delta R$ represents the surface area of a disc annulus of width
$\Delta R$ at a given distance $R$ from the black hole. The energy
equation is then 
\begin{equation}
\frac{dq}{dt}=A\left(Q^+-Q^-\right).
\end{equation}

The source terms on the RHS of the energy equation (\ref{eq:energy})
are given by the viscous dissipation of the disc per unit annulus area, 
\begin{equation}
  \label{eq:viscdisp}
  Q^+=\frac{9}{8}\nu_\textrm{t} \Sigma \frac{GM}{R^3},
\end{equation}
and radiative losses, $Q^-$, which we take to be of the form
\begin{equation}
  \label{eq:radloss}
  Q^-=\frac{4\sigma}{3\tau}T^4,
\end{equation}
where the optical depth $\tau$ is given in terms of the opacity
$\kappa_\textrm{R}$ by 
\begin{equation}
\tau=\frac{1}{2}\Sigma
\kappa_\textrm{R}(\rho,T). 
\end{equation}
To ensure the consistency of the model we need disc to be optically
thick in the vertical direction, i.e. $\tau
\gg 1$.  

We have written the numerical scheme so that internal energy $e$ is
conserved to machine accuracy, apart from losses on the inner and
outer boundary and local source terms (i.e. $PdV$). The boundary
condition at the outer disc edge for the energy equation is taken to
be divergence free, i.e. energy is being put in the last cell as it is
lost to the next innermost cell. At the inner boundary energy is lost,
assumed captured by the black hole.

\subsection{Material functions}

\subsubsection{Equation of state} \label{sect:eos}

The total pressure $P$ is given by the sum of gas and radiation
pressure (since $\tau\gg 1$)
\begin{equation}
  \label{eq:pressure}
  P=\rho \frac{k_\textrm{B} T}{\mu m_\textrm{p}}+\frac{4\sigma}{3c} T^4,
\end{equation}
while the specific internal energy $U$ of the mixture of monatomic gas
\& radiation is 
\begin{equation}
  \label{eq:u} U= \frac{3}{2}\frac{k_\textrm{B} T}{\mu
  m_\textrm{p}}+\frac{4\sigma}{c\rho} T^4.
\end{equation}
The sound speed $c_s$ can be calculated using
\begin{equation}
  \label{eq:cs}
  P=\gamma\rho c_s^2, 
\end{equation}
where the adiabatic index $\gamma$ is given by
\begin{equation}
\gamma=\left(\frac{\partial \log
      P}{\partial \log \rho}\right)_S =
  \frac{16-12\beta_P-\frac{3}{2}\beta_P^2}{12-\frac{21}{2}\beta_P},
\end{equation}
and is a slowly varying function of $\beta_P$, which ranges from
$\frac{4}{3}$ to $\frac{5}{3}$ as $\beta_P$ ranges from $0$ to $ 1$,
where $\beta_P$ is the ratio of gas pressure to total pressure.

\subsubsection{Viscosity} \label{sect:viscosity} 

We use the standard $\alpha$-viscosity prescription put forward by
\cite{1973A&A....24..337S} and take the  the $(r\phi)$-component of the
stress tensor to be $t_{r\phi}=-\alpha P$. Relating this to the
rate-of-strain tensor, through the viscosity $\rho\nu_{\rm t}$
\citep[e.g.][]{2001NewAR..45..663B} so that $t_{r\phi}=-\frac{3}{2} 
\nu_{\rm t}\rho \Omega_\textrm{K} $ we see that the kinematic viscosity is
related to $\alpha$ by the equation  
\begin{equation}
  \label{eq:viscosity} 
\nu_{\rm t} = \frac{2}{3}\alpha \frac{P}{\rho \Omega_\textrm{K}}.
\end{equation}

This viscosity prescription (eq.~\ref{eq:viscosity}), using
Eqns.~(\ref{eq:hydrostat}), (\ref{eq:sigma}) and~(\ref{eq:cs}), can
also be written in the form
\begin{equation}
\nu_{\rm t} = \frac{2}{3}\alpha \sqrt{\gamma} c_s H.
\end{equation}
If we envisage the viscosity $\nu_{\rm t}$ in physical terms as the
product of a characteristic turbulent length and turbulent velocity,
$l_\textrm{t}$ and $V_\textrm{t}$, and our disc is assumed to be
geometrically thin, we can associate the characteristic velocity and
length scale with a fraction of the local scale height and of the
local sound speed, respectively. In order that the turbulence remains
subsonic we see that the factor $2/3\alpha \sqrt{\gamma}$ needs to be
smaller than, or of order, unity.

The viscous torque (\ref{eq:torque}) using the viscosity prescription (\ref{eq:viscosity}) can be written as
\begin{equation}
  \label{eq:torque2}
  G=-4\pi\alpha P H R^2\;.
\end{equation}

\subsubsection{Opacity} \label{sect:opacity}

We take the Rosseland mean opacity, $\kappa_\textrm{R}$, tabulated by
the OPAL opacity
project\citep{1992ApJ...401..361R,1996ApJ...464..943I} for the solar
composition of \citet{GN93}. We use the $X=0.7$ set of their opacity
tables ($X$ is the hydrogen mass fraction of the matter). Without further
notice we fix the metallicity to be $Z=0.02$ (solar metallicity).

\subsection{Numerical Grid and calculation procedure}
\label{sect:grid}

We use a one-dimensional grid in radius with N points ranging from
$R_\textrm{in}=3R_\textrm{S}$ to $R_\textrm{out}$. We solve the
time-dependent equations (\ref{eq:angmom1}), (\ref{eq:kg1}) and
(\ref{eq:energy}) using the mass $m$ and internal energy $e$ of a disc
semi-annulus as dependent variables.

In order to allow simple application of the concepts of variable
dynamo-driven angular momentum presented by
\citet{2004MNRAS.348..111K}, it is necessary for the widths ($\Delta
R$) of the disc annuli to be comparable to the local disc thickness
$H$. Within the context of the one-zone approximation, this is a
reasonable assumption, since in this case it makes little physical
sense to try to attribute physical reality to attempts at resolving
structures significantly smaller than the scale height $H$ of the
disc. \footnote{\citet{1998MNRAS.298.1048H} demonstrate that formal
mathematical convergence of the numerical scheme may require a grid
resolution much less than the disc thickness. We note here that
mathematical convergence does not necessarily imply more accurate
modelling of physical reality.}

\begin{figure}
  \includegraphics[width=0.35\textwidth,angle=-90]{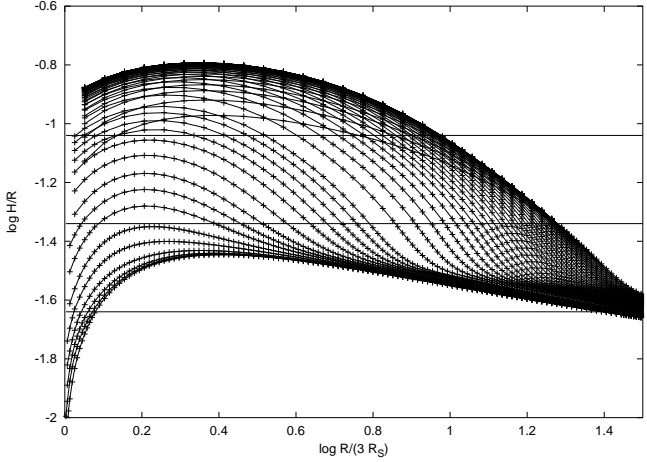}

\caption{The refinement of the numerical grid. The Figure shows log
$H/R$ versus log $R$ as it changes in a typical run. Every time the
local scale height $H$ exceeds(falls short of) twice(half) the local
grid cell width, the resolution is reduced(increased). Due do the
binarity requirement of the grid (see text) the same grid points are
restored after repeated refinement/coarsening. } \label{fig:grid}
\end{figure}

While $R_\textrm{in}$ and $R_\textrm{out}$ are kept constant, the
number of grid-points is time-dependent and the number of grid points
is adjusted according to the following prescription (see
Fig.~\ref{fig:grid}). We remove/insert grid points if the local radial
extent of the annulus is larger/smaller than $2H$ and $0.5H$,
respectively. Grid points are inserted by conserving disc mass,
angular momentum and energy $e$.  In typical runs, $N$ ranges between
a few dozens and a few hundred points.

The setup of the grid is implemented as follows. At the start of the
calculation there are 2 points, at $R_\textrm{in}$ and
$R_\textrm{out}$. Then we calculate the scale height
$H(R_\textrm{out})$ and compare it with the interval
$R_\textrm{out}-R_\textrm{in}$. If $R_\textrm{out}-R_\textrm{in}> 2
H(R_\textrm{out})$, we put a new grid point at
$R_\textrm{new}=\sqrt{R_\textrm{out}R_\textrm{in}}$ and calculate the
scale height $H(R_\textrm{new})$. This procedure is recursively
applied on both intervals. The setup is complete if there is no need
to refinement any more.

From a numerical point of view it is beneficial in preventing
numerical mixing to use a scheme which retains the structure of the
grid cells fixed as far as possible.  Thus, in order to maintain a
binary structure of the grid (yielding the same coarse interval at the
same place after several refinements), we assign each grid point with
a number $N$, initially $N(R_\textrm{in})=0$ and
$N(R_\textrm{out})=1$. As a refinement occurs, the new grid point gets
the mean number of the adjacent grid points. If this number is not an
integer, all $N$ are multiplied by 2.  The condition of grid
coarsening is accompanied by the condition for keeping the grid
binary, i.e. we only remove a point if the
$N(R_\textrm{outer})-N(R_\textrm{remove}) =
N(R_\textrm{remove})-N(R_\textrm{inner})$ and $N(R_\textrm{inner})$
has to be an integer multiple of $N(R_\textrm{outer}) -
N(R_\textrm{inner})$. This binarity of the grid ensures that, for
example, the magnetic field is never mixed numerically across the
entire grid, i.e. subsequent refinement and coarsening restores the
same grid cell again.

A typical coarsening/refinement can be seen in
Fig.~\ref{fig:grid}. The $\Delta (\log R)$ correlate very well with
the $H/R$ ratio since $\Delta (\log R)=\log ((R+\Delta R)/R)=\log
(1+\Delta R/R)\approx \Delta R/R = H/R$ as required. Every $\Delta
\log H/R=0.3$ dex, i.e. at $\log H/R\approx-1,-1.3$ and $-1.6$ the
resolution changes, as $H$ de-/increases by a factor of 2. 

\subsection{Advection} \label{sect:advection}

It is often the case \citep[see, for example, the discussion
in][]{1986MNRAS.221..169P} that the advective terms in the energy
equation (i.e. those terms containing the radial velocity $u_R$ on the
RHS of Equation~\ref{eq:energy}) are omitted. However, it was realised
early that even in the case of dwarf nova discs these terms can make a
difference \citep{1983MNRAS.205..359F}. And more recently,
\citet{1988ApJ...332..646A} drew attention to the fact that the radial
advection of heat can play an important role in the local energy
balance in the disc, especially at high accretion rates close to the
black hole, and can, in particular, have a stabilising effect on the
disc in that region.  To illustrate this, we note that the terms
responsible for the radial advection of energy can be written in terms
of the radial entropy gradient as if they correspond to a local
additional heat loss term \citep[e.g.][]{1982AcA....32....1M,2001NewAR..45..663B}, in the form
\begin{equation}
Q_\textrm{ad}=\frac{\dot M T}{2\pi R}\frac{\partial S}{\partial
  R}.
\end{equation}
The radial entropy gradient can be expanded in terms of pressure and
radial density and temperature gradients. In doing so we obtain

\begin{equation}
Q_\textrm{ad}=\frac{\dot M }{2\pi R^2}\frac{P}{\rho}\chi_\textrm{ad} ,
\end{equation}
where the dimensionless measure of the strength of the effect,
$\chi_\textrm{ad}$, includes the radial derivatives. Usually,
$\chi_\textrm{ad}(R)$ is a slowly varying function close to
unity. However, although in the gas pressure dominated regime this
value is indeed close to unity, in the radiation pressure dominated
regime $\chi_\textrm{ad}$ can be as large as 10,

Using the expressions for the internal energy and pressure in Appendix
\ref{sect:appenergy}, we can express $\chi_\textrm{ad}$ in terms of the
radial derivatives of $\rho$ and $T$ for a stationary disc 
\begin{equation}
\chi_\textrm{ad} =
-\left(12-\frac{21}{2}\beta_P\right)\frac{\partial \log
  T}{\partial \log R}+\left(4-3\beta_P\right)\frac{\partial
  \log \rho}{\partial \log R}. 
\end{equation}

\section{Stationary solutions and stability} \label{sect:stationary}

Before investigating the time dependent disc behaviour, and before
adding in the complications of stochastically driven accretion, we
first investigate steady disc solutions, and the conditions for
stability.

\subsection{Stationary solutions}
\label{sect:statsolns}

In the stationary case, with accretion rate $\dot{M} = $ const., we
have $\partial/\partial t=0$ and thus
the conservation of angular momentum (Eqn.~\ref{eq:angmom1}) becomes
\begin{equation}
  \dot M f = 3\pi \nu_\textrm{t} \Sigma, 
\end{equation}
conservation of energy (\ref{eq:energy}) becomes
\begin{equation}
 Q^+ -  Q^- - Q_\textrm{ad} = 0,\label{eq:statenergy}
\end{equation}
and vertical hydrostatic equilibrium (\ref{eq:hydrostat}) remains
\begin{equation}
 P  =  \frac{GM}{4\rho R^3}\Sigma^2,
\end{equation}
where the function $f(R)=1-(R/R_\textrm{in})^\frac{1}{2}$ ensures
compliance with the inner boundary condition. For a given accretion
rate, and at a particular radius, these three equations can be written
as two equations, depending on pressure $P$, density $\rho$ and
temperature $T$. The energy equation yields

\begin{equation}
\label{eq:energystat}
 \frac{\dot M}{2\pi
  R^2}\left(\frac{3}{4}\Omega_\textrm{K}^2R^2
  f-\frac{P}{\rho}\chi_\textrm{ad}\right)-\frac{16\pi \sigma T^4
  \alpha P}{3\kappa_\textrm{R}\rho \Omega_\textrm{K}\dot M f} = 0,
\end{equation}
and hydrostatic equilibrium yields
\begin{equation}
  \label{eq:energystat2} P^3-\frac{\Omega_\textrm{K}^4\dot M^2 f^2
 \rho}{16\pi^2\alpha^2} = 0.
\end{equation}
Then for a given accretion rate, and at a particular radius, these two
equations, together with the equation of state
(Equation~\ref{eq:pressure}) can be solved for $P$, $\rho$ and $T$.

We solve these equations with a 2D nested-intervals method as
described in \citet{2005MNRAS.356....1M}. In general we consider the
conditions at a particular radius $R$, for various values of the
accretion rate $\dot M$. For illustrative purposes we also consider
the solutions for three different particular values of the strength of
the adiabatic heat flux $\chi_\textrm{ad}$, and for various values of
the viscosity parameter $\alpha$.

We present our results in terms of the Eddington luminosity and the
corresponding Eddington accretion rate.  In terms of the Eddington
luminosity, $L_{\rm Edd} = 4 \pi GM m_{\rm p} c/\sigma_{\rm T}$ we may
define a corresponding Eddington accretion rate by
\begin{equation}
  \label{eq:ledd}
L_\textrm{Edd}=\eta_\textrm{K}\dot M_\textrm{Edd} c^2,
\end{equation}
where $\eta_K$ is the efficiency. For a typical value of $\eta_K =
1/12$ (see below) we find in numerical terms
\begin{equation}
\label{eq:edd} 
\dot M_\textrm{Edd}= 2.68\cdot10^{-7} \cdot
\left(\frac{M}{10 M_\odot}\right) \textrm{M}_\odot \mbox{yr}^{-1}.
\end{equation}

\begin{figure}
  \centering
    \includegraphics[angle=-90,width=0.49\textwidth]{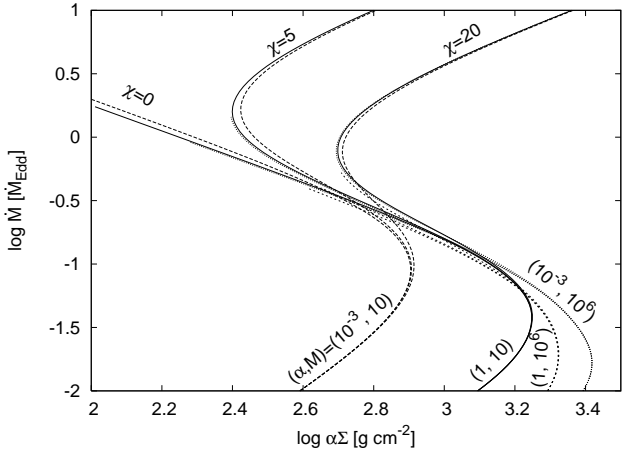}
    \caption{Local equilibrium solutions at $R=30 R_S$ for a range of
    accretion rates. We consider solutions for different values of
    viscosity parameter $\alpha$, the black hole mass $M$ and the
    strength of radial advection, $\chi_\textrm{ad}$. Parts with
    negative $\dot M-\Sigma$ slope are thermally and viscously
    unstable. Note the stabilising effect of advection (when
    $\chi_\textrm{ad}\neq 0$).}  \label{fig:scurve}
\end{figure}

\subsection{S-Curves}
\label{sect:scurves}

In Fig.~\ref{fig:scurve} we plot the relationship between accretion
rate $\dot{M}$ and $\alpha$ times disc surface density $\Sigma$ for
different values of radially constant $\alpha$ and for different
values of $\chi_\textrm{ad}$ and of the central mass $M$.  

These plots give us direct information about the stability of the
steady-state discs. There are two types of instability which these
discs can be subject to -- viscous
\citep[e.g.][]{1974ApJ...187L...1L} and thermal
\citep[]{1976MNRAS.177...65P}. We show in Appendix~\ref{sect:stabanalysis} that the
stability criterion in both cases is the same. The discs are unstable
when
\begin{equation}
\left(\frac{\partial \log \dot M}{\partial \log
\Sigma}\right)_{P,T} < 0.
\end{equation}

From Figure~\ref{fig:scurve} we see
that in the absence of heat advection the discs become unstable
(i.e. the $\dot{M}(\Sigma)$ curves have negative gradient) as the
accretion rates approach the Eddington limit. But, as pointed out by
\citet{1988ApJ...332..646A}, advection of heat (i.e. non-zero $\chi_\textrm
{ad}$) provides stability. For all combinations of $\alpha$ and $M$
considered, we see
that within the physical range for $\chi_\textrm
{ad}$ there are always
values of the accretion rate for which the discs are unstable, whereas
for higher black hole mass $M$, higher $\alpha$ and lower
$\chi_\textrm{ad}$ the range of accretion rates we get instability
becomes larger. 

It is
well known \citep[see, for example,][]{2001NewAR..45..449L} that the instability causes
the disc to undergo limit-cycle behaviour. We investigate this further
below.

\subsection{Dependency on black hole mass and metallicity}

For black hole masses typical of X-ray binaries ($M\approx
10$ M$_\odot$), characteristic mid plane temperatures close to the
black hole are of the order of $10^7$ K. 

Since $\dot M \Omega_K^2\propto T^4/\tau_R$ (cf. eq.~\ref{eq:statenergy}) in the absence of
advection, we get in scaled units ($R$ in Schwarzschild Radii, $\dot M$ in
units of Eddington accretion rate) and roughly constant optical depth
that $T\propto M^{-1/4}$. For a black hole mass of $10^6$ M$\odot$ we 
then expect the mid plane temperature of the disc to be of the order of 
a few times $10^5 K$.  

\begin{figure}
  \centering
    \includegraphics[angle=-90,width=0.49\textwidth]{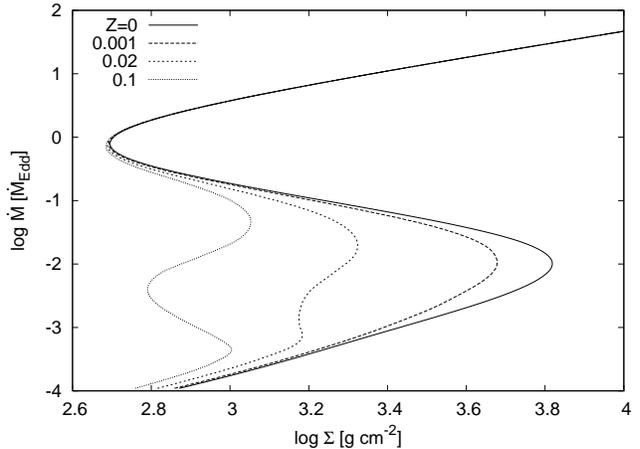} 
  \caption{Local equilibrium solution at $R=30R_S$ for $M=10^6$ M$_\odot$,
    $\alpha=1$, $\chi_{ad}=20$ and different metallicities
    $Z=0,0.001,0.02$(solar) and $Z=0.1$. The hydrogen content is
    fixed at $X=0.7$. }
  \label{fig:scurve3}
\end{figure}

For $10^6$ $M_\odot$ the disc temperature reaches $T\approx 10^{5.4}$ K for reasonable
accretion rates. Then the disc shows an additional instability caused
by the ''Z-Bump'' \citep[cf. ][]{1994MNRAS.266..805S} in the opacity for
a metallicity larger than $Z=0.02$. This so called Z-Bump is also 
responsible for pulsations in
$\beta$ Cep stars \citep{1982ApJ...260L..87S,1992MNRAS.255P...1K}. We
show a S-curve presenting this instability in
Fig.~\ref{fig:scurve3}. It is evident that the disc for this
instability locally can jump from accretion rates of $10^{-4}\dots 10^{-3}$
to $10^{-1} \dot M_\textrm{Edd}$.

Generally, the disc becomes more unstable with increasing mass of the
black hole, i.e. the lower turning point in the ''S-curve'' moves to
lower accretion rates.

\subsection{$H/R$ in the presence of advection}

For the approximations used in this paper to be valid we require that
the disc thickness be small compared to the radius. We note here that
the addition of a local heat sink in the guise of advection implies
that the disc is typically thinner than it would be if all the energy
generated locally were radiated at the same radius.  In the stationary
case, when advection of heat strongly dominates radiative losses (so
that $Q_\textrm{ad}\gg Q^-$), Equation~(\ref{eq:energystat}) becomes
\begin{equation}
\frac{3}{4}\left(\Omega_\textrm{K} R\right)^2 f -
\frac{P}{\rho}\chi_\textrm{ad}=0.
\end{equation}
Then using hydrostatic equilibrium (Equation~\ref{eq:hydrostat})
we find that
\begin{equation}
\label{eq:advdom}
\frac{H}{R}=\sqrt{\frac{3f}{4\chi_\textrm{ad}}}.
\end{equation}

Thus, for example, taking $\chi_\textrm{ad}=10$ as a representative
value, and with $f~=~\frac{1}{2}$ (corresponding to $R\approx 10
R_\textrm{S}$) we find $H/R \approx 0.2$. Indeed, we see that as long as
$\chi_\textrm{ad}> 4/3$, $H/R$ is smaller than unity.

\subsection{Keplerian rotation} 
\label{sect:keplerian}

For similar reasons, if advective heat flow is important, the disc
angular velocity stays close to Keplerian. If we include the radial
pressure gradient in the radial momentum balance equation, we find
that the disc angular velocity $\Omega$ is given by
\begin{equation}
\label{eq:radmom}
\Omega^2=\frac{GM}{R^3}\left(1+\frac{PR}{GM\rho}\xi_P\right),
\end{equation}
where 
\begin{equation}
-\xi_P=\frac{\partial \log P}{\partial \log R}.
\end{equation}
This may be written as
\begin{equation}
\xi_P = \beta_P\frac{\partial \log \rho}{\partial \log
R}+\left(4-3\beta_P\right)\frac{\partial \log T}{\partial \log R}.
\end{equation}

The dominant term in the radiation pressure dominated domain
($\beta_P=0$) is, of course, the temperature gradient, and a
representative value is $\xi_P\approx 2$.

In (\ref{eq:radmom}) we treat the extra term in the brackets as small
deviation. Thus we can replace $GM/R$ by $\left(\Omega_\textrm{K}
  R\right)^2$ and with the hydrostatic equilibrium
(\ref{eq:hydrostat}) we get
\begin{equation}
\Omega^2=\Omega_\textrm{K}^2\left(1+\left(\frac{H}{R}\right)^2\xi_P\right).
\end{equation}
When advective heat transport is strongly dominant we can use
Equation~(\ref{eq:advdom}) to deduce that
\begin{equation}
\Omega^2 =
\Omega_\textrm{K}^2\left(1+\frac{3f\xi_P}{4\chi_\textrm{ad}}\right).
\end{equation}
For small deviation ($\Omega-\Omega_\textrm{K}\ll\Omega_\textrm{K}$)
this gives approximately
\begin{equation}
\frac{\Omega}{\Omega_\textrm{K}}-1=\frac{3f\xi_P}{8\chi_\textrm{ad}}.
\end{equation}
Thus, using representative values, the deviation from Keplerian
rotation when advective heat transport is strong never exceeds around
4 per cent. The assumption of Keplerian rotation is therefore a reasonable one.

\section{Time-dependence - Results} 
\label{sect:timedep}

We have seen above that even in the absence of stochastic magnetic
phenomena our accretion disc models are expected to display
time-dependent behaviour for some ranges of the parameters.  We have
carried out a number of simulations for different values of $\alpha$,
different external (or mean) accretion rates and different black hole
masses. 

In this section, for reasons of computational time, we use a logarithmically
equidistant grid with $N=500$ grid points without refinement and
coarsening.
We solve the equations with
an implicit solver. Comparison with the standard explicit solver and
the refinement and coarsening as described in Sect.~\ref{sect:grid}
does not show significant deviations. However the computing time
decreased drastically and thus enabled us to carry out a parameter study
concerning the global stability of these discs with respect to the
instabilities discussed above. 

\subsection{General behaviour}

Our time-dependent models depend on the mass of the black hole, the
external accretion rate and the value of the viscosity
parameter $\alpha$.

As initial conditions we set up a stationary disc, assuming $Q^+=Q^-$
and neglecting advection. Depending on the accretion rate and the
viscosity parameter, the disc either continues to stay in this
stationary state, or, if advection of heat is important, the inner
disc re-adjusts itself to allow for the advection. As we see from
Figure~\ref{fig:scurve} if the viscosity is high enough, we expect
limit cycles to appear with the inner disc oscillating between a hot
state with high accretion rate and a cool state with lower accretion
rate (see Section~\ref{sect:scurves}).

If the inner disc is such that the mean accretion rate produces
instability, the inner disc spends its time trying to jump between the
two stable branches of the S-curve. While it is on the upper branch,
with advection important, the disc is steadily depleted. The
surface density decreases while the accretion rate is higher than
average. As the inner disc is depleted, some matter is transported
outward in a heating front. The inner disc is radiation pressure
dominated and the accretion rate there is radially constant. Matter
then piles up ahead of this front and subsequently the inner disc
begins to starve and the accretion rate decreases.  When the front
reaches the stable, gas pressure dominated region of the disc, the
front slows down and the inner disc cools, becoming more gas pressure
dominated, as the supply of matter is interrupted. The dissipated
energy is advected outwards and the accretion rate in the inner disc
decreases, varying radially as $\dot M\propto R^2$. Then inner disc
reheats, the accretion rate increases inwards, the pileup is accreted
and eventually the cycle restarts with the inner disc again becoming
depleted.

\begin{figure}
  \centering
  \includegraphics[angle=-90,width=0.5\textwidth]{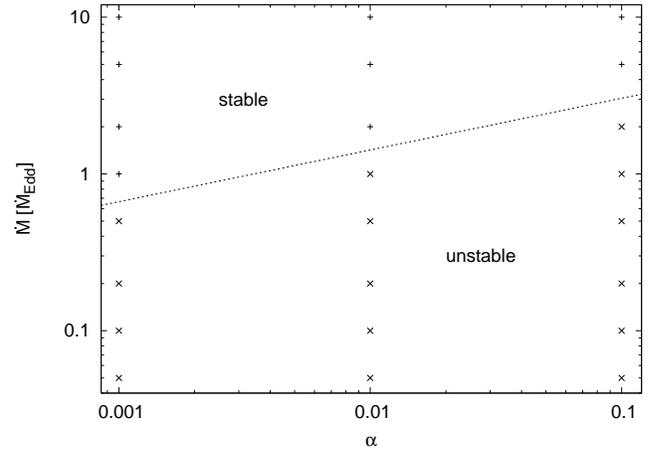}
  \caption{$\dot M$ versus $\alpha$ showing the systems with 
    limit cycles (to the lower right of each
  line) and without (upper left) for a 10 M$_\odot$ black hole. The line is only a rough indication
  since the grid in $\left(\dot M,\alpha \right)$ is very coarse. The
  points, for which actually calculations have been done, are
  indicated by the crosses.}  \label{fig:dm-m}
\end{figure}

\subsection{Influence of $\alpha$ and the black hole mass on the stability}
\label{sect:nolimit}

The parameters for the grid of models with constant $\alpha$ discs we
ran are shown in Figure~\ref{fig:dm-m} for a 10 M$\odot$ black hole. 
Also shown in
Figure~\ref{fig:dm-m} are those parameter values for which limit-cycle
behaviour is found. The higher the viscosity
parameter, the more unstable the disc becomes since then advection can
only stabilise for higher accretion rates.

For higher black hole masses, there are several complications.
First, the 
radiation pressure dominated zone
and hence the \citet{1974ApJ...187L...1L} unstable region is
larger. For $10^7$ M$_\odot$ this zone is so large (more than 1000
Schwarzschild radii), that there is likely to be an interaction
between this instability and the usual dwarf nova instability, caused
by the opacity drop at hydrogen ionisation. This has yet to be
explored \citep[cf.][]{1988MNRAS.235..881C,1989ApJ...338...32C}. 

Second, as discussed previously, there is a ''Z-Bump'' instability
present which makes the results extremely dependent on the chosen
metal content for the opacity. 

Finally the computational effort
becomes increasingly larger. The disc in the
\citet{1974ApJ...187L...1L} unstable region oscillates between a thick
disc and a more and more ultra-thin disc ($H/R\approx 10^{-4}\dots
  10^{-3}$) which puts strong limits on the time-step. 

These restrictions led us to concentrate on the 10 $M_\odot$ black
hole case and to leave the higher
mass case for further investigations. We plot a sample lightcurve and
a power density spectrum for a $10^6$ $M_\odot$ black hole in
Fig.~\ref{fig:pspectra6}. The two different instabilities \citep[][and the
''Z-Bump'' instability ]{1974ApJ...187L...1L} operate on different
timescales and different amplitudes. The \citet{1974ApJ...187L...1L}
instability leads to the huge outbursts in luminosity, while the small
fluctuations are due to the ''Z-Bump'' instability. For the example
chosen this instability affects only the region around 100 Schwarzschild
radii. The instability then only appears as small fluctuations in the
total disc luminosity. Note that for computational feasibility we set
the number of grid points artificially low. In order to fulfil our resolution criterion
($\Delta R\approx H$, cf. Sect.~\ref{sect:grid}), we would have needed
$N\approx 3000$ points to properly resolve the inner disc in the
low-$\dot M$ state. Thus the lightcurve is only indicative of the
complications that can occur.

\subsection{Effect of advection on the observed accretion rate}

By definition the radiative luminosity of the disc is 
\begin{equation}
\label{eq:lrad}
  L_{\rm rad} = 2\pi\int_{R_\textrm{in}}^{R_\textrm{out}} Q^- R dR. 
\end{equation}
For an infinitely extended, steady standard disc
($R_\textrm{out}\to\infty$, no radial advection of heat),
the energy dissipated in the disc by viscous processes (using our
inner 'black hole' radius $R_\textrm{in}=3R_\textrm{S}=6GM/c^2$) is 
\begin{equation}
  \label{eq:totradloss} L_{\rm diss} = \frac{1}{2} \frac{G M \dot
  M}{R_\textrm{in}}=\frac{1}{12}\dot M c^2.
\end{equation}
\begin{figure}
  \centering
  \includegraphics[angle=-90,width=0.45\textwidth]{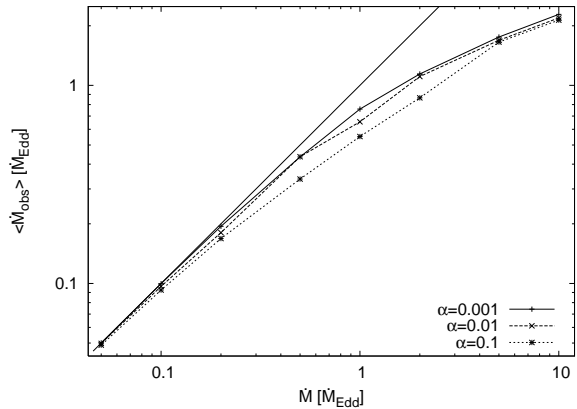}
  \caption{The time-averaged 'observed' accretion rate, $\dot{M}_{\rm
    obs}$, deduced from the radiative luminosity $L_\textrm{rad}$, is
    plotted versus the external (actual) accretion rate $\dot M$ for a
    10 $M_\odot$ black hole for
    different values of $\alpha$. For high actual accretion rates,
    some of the accretion energy is advected into the black hole. }
    \label{fig:eddington-ratio}
\end{figure}
Thus the efficiency $\eta_\textrm{K}$ of our Newtonian, standard black hole
accretion disc in Keplerian rotation extending from
$R_\textrm{in}=3R_\textrm{S}$ to infinity in converting rest mass
energy into radiation is $\eta_\textrm{K}=\frac{1}{12}$.

In the absence of advective heat flow, for an assumed value of the
efficiency $\eta_K$ a measurement of the observed radiative flux,
i.e. $L_{\rm rad}$, yields an estimate of the accretion rate by
setting $L_{\rm rad} = L_{\rm diss}$.
However, when advection of heat is important, we expect deviations
from the standard $L_\textrm{rad}\propto \dot M$ relation. In
particular, since some of the dissipated heat is now advected directly
into the black hole, we expect the observed radiated flux to give an
underestimate of the actual accretion rate.  From the numerical
simulations, we calculate the radiative luminosity $L_{\rm rad}(t)$ at a
given time, and
deduce from an 'observed' accretion rate given by
\begin{equation}
\dot{M}_{\rm obs}(t) = \frac{L_{\rm rad}(t)}{\eta_K c^2}\;,
\end{equation}
or the time-average of this
\begin{equation}
\left<\dot{M}_{\rm obs}\right> = \frac{1}{\eta_K c^2 T}\int^T L_{\rm rad}(t)\; dt\;.
\end{equation}
In Fig.~\ref{fig:eddington-ratio} we plot $<\dot{M}_{\rm obs}>$ versus
the actual accretion rate $\dot{M}$, both in units of the Eddington
accretion rate. If the disc model at the given set of parameters
shows limit-cycle behaviour, we calculate the time-averaged value of
$L_{\rm rad}$ for an integer number of complete limit cycles.

For a 10 M$_\odot$ black hole at low
accretion rates, we find as expected $<\dot{M}_{\rm obs}> =
\dot{M}$. With increasing accretion rate, however, the advection comes
into play and the ''observationally calculated'' accretion rate
deviates from the actual $\dot M$. For a 10 M$_\odot$ black hole this
relation is practically independent of $\alpha$. This deviation may be
related to the deviations from the $L_\textrm{rad}\propto T_\textrm{in}^4$
relationship recently reported by \citet{2004ApJ...601..428K} and
\citet{abe-2005-}. They find the so called 'apparently standard'
regime in observations of XTE J1550-564 and 4U 1630-47, where the disc
luminosity is proportional to $T_\textrm{in}^2$ and attribute this to
effects of the radial advection of energy.

For the $10^6$  M$_\odot$ black hole (Fig.~\ref{fig:pspectra6}). There are departures from $<\dot{M}_{\rm obs}> =
\dot{M}$ . While the disc is
being fed with an accretion rate of $\dot M=0.1\dot M_\textrm{Edd}$, $<\dot
M_\textrm{obs}>$ is  only 44 \% of the actual accretion rate,
while the time average for the low-$\dot M$ state only is as low as
1.5 \% of the actual accretion rate. This is a
result of the strong outburst behaviour of these discs compared to
the low-mass case. Most of the mass previously stored in the outer
disc is pushed through the inner disc
in the high-$\dot M$ state. Then, however, most of the energy created
by viscous dissipation is advected into the black hole. Since the
outbursts are short compared to the complete limit cycle, the
efficiency is fairly low.

\subsection{Power density spectra}

We calculate lightcurves at equidistant time points. Suppose we have
a lightcurve covering the time $T$ with $N$ points.  We calculate
power density spectra using the canonical normalisation to get the
Power $P_\nu$ in units of (rms/mean)$^2$/Hz. Integrating $P_\nu d\nu$
gives the integrated fractional rms. We compute the FFT of the lightcurve
and then define $P_\nu$ by \citep{1988SSRv...46..273L}
\begin{equation}
  \label{eq:pds}
  P_\nu=\frac{2\left|a_\nu\right|^2}{\bar L^2 T}\left(\frac{T}{N}\right)^2
\end{equation}
where $\bar L$ is the mean luminosity of the lightcurve and $a_\nu$ the
non-normalised Fourier coefficient. The $P_\nu$'s are binned into
logarithmically equidistant points. For the FFT we use the routines
from the FFTW library\footnote{http://www.fftw.org/}. 

The integrated fractional rms is given in percent
\begin{equation}
  \label{eq:frms}
  r=100 \% \cdot \sqrt{\int P_\nu d\nu } 
\end{equation}

Using Parseval's theorem, it can be shown, that $r$ indeed corresponds
to the rms/mean of the luminosity fluctuations
\begin{equation}
r=100 \% \cdot \sqrt{\frac{1}{N}\sum_{i=1}^N \left(\frac{L_i}{\bar L}-1\right)^2}
\end{equation}
where $L_i$ is the luminosity given at equidistant time intervals and
$\bar L$ the time-average of the total light curve. Unless otherwise
stated, the last formula is used to calculate the integrated fractional rms.

\subsection{Period of limit cycle}

\begin{figure*}
  \centering
  \begin{tabular}{cc}
   \includegraphics[width=0.33\textwidth,angle=-90]{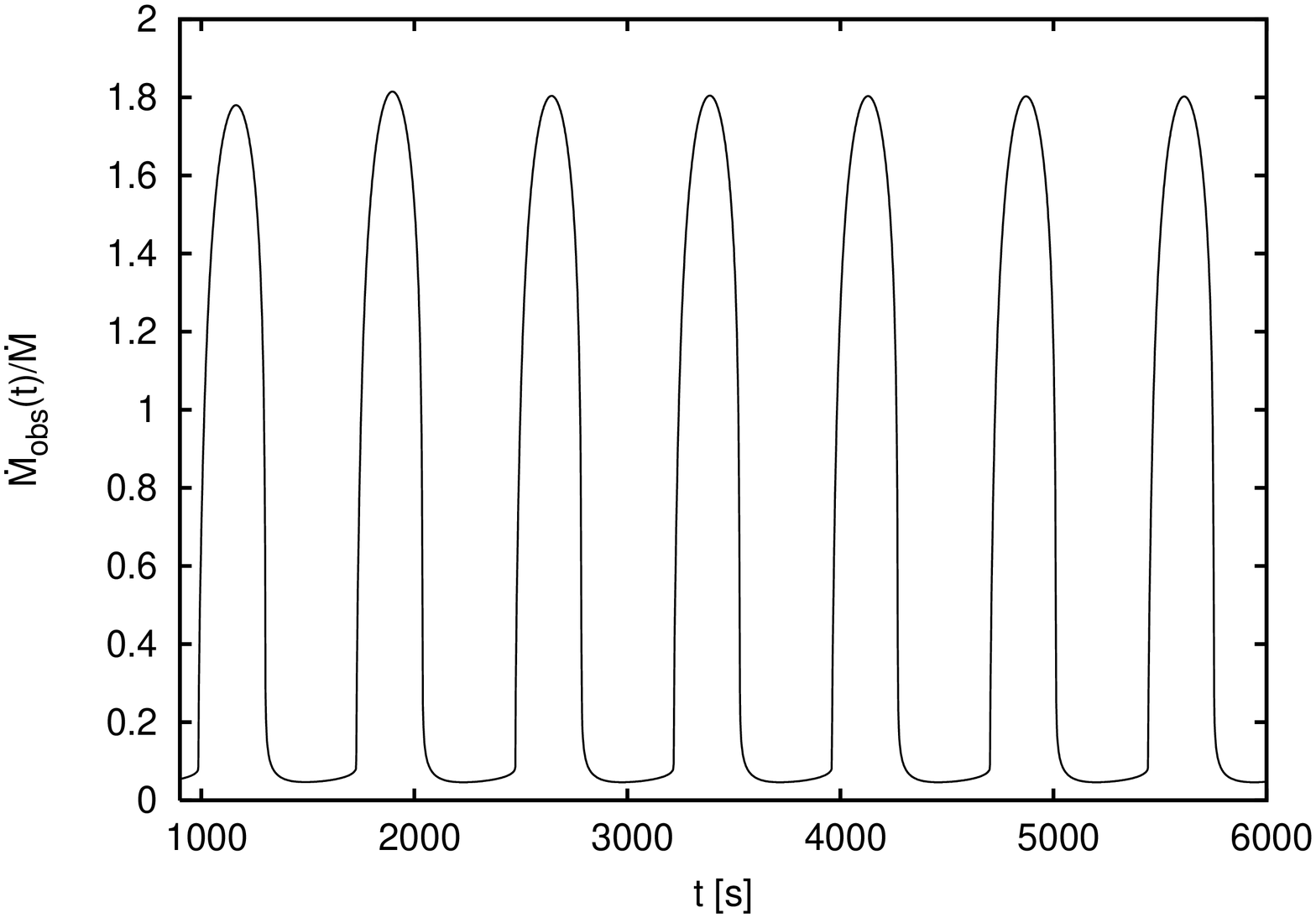} & \includegraphics[width=0.33\textwidth,angle=-90]{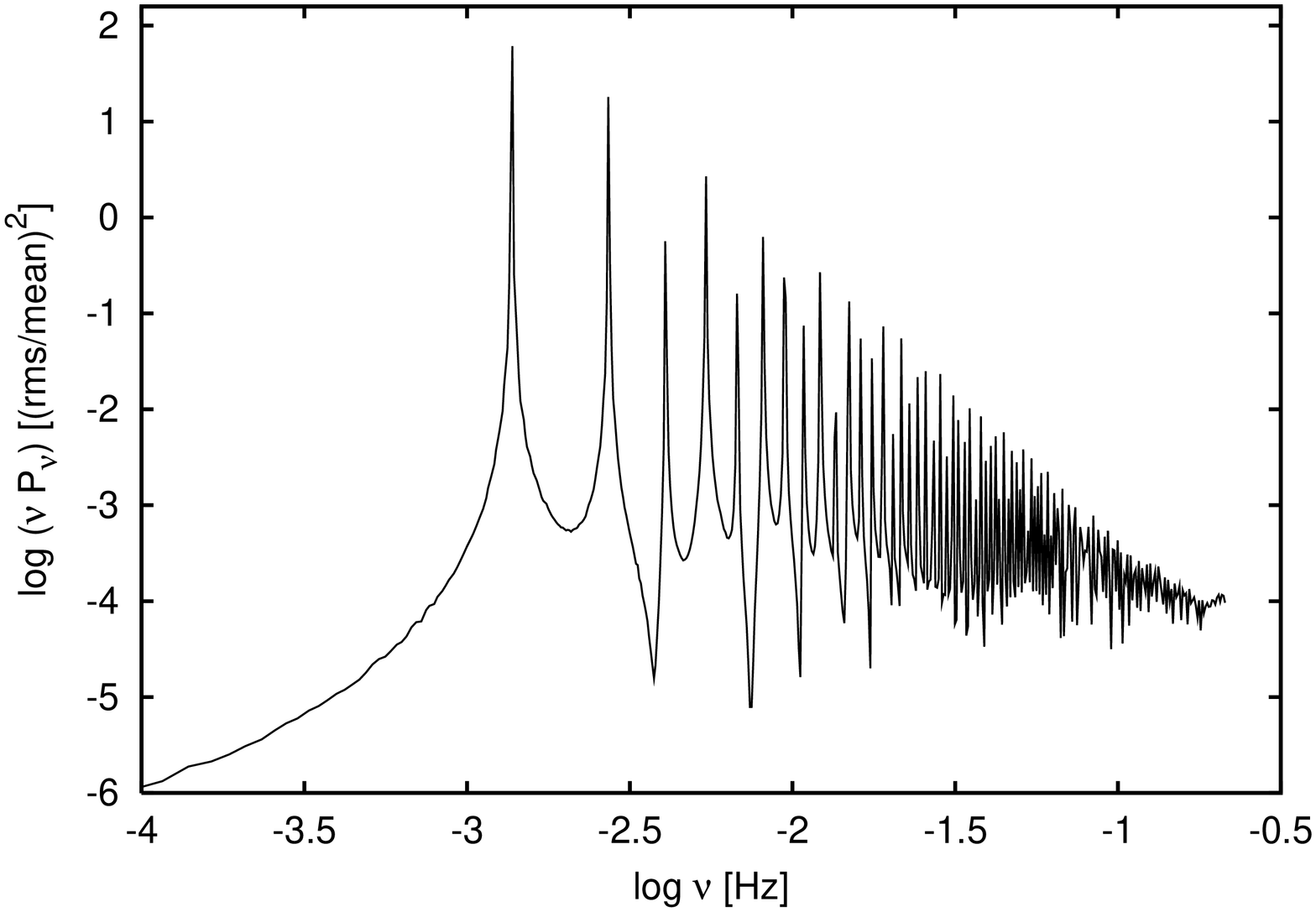}
  \end{tabular}
  
  \caption{{\bf (left)} Lightcurve (detail) and {\bf (right)} Power density spectrum for an accretion disc around a 10
  M$_\odot$ black hole accreting at 0.5 $\dot
  M_\textrm{Edd}$. $\alpha=0.1$. The FFT data are binned in
logarithmically equispaced bins of width $\Delta \nu/\nu=0.005$. } 
  \label{fig:pspectra}
\end{figure*}

\begin{figure}
  \centering
  
   \includegraphics[width=0.33\textwidth,angle=-90]{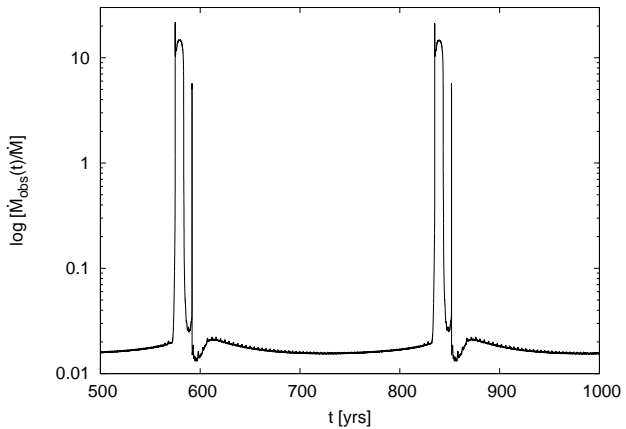}

  \caption{Lightcurve for an accretion disc around a $10^6$
  M$_\odot$ black hole accreting at 0.1 $\dot
  M_\textrm{Edd}$. $\alpha=0.1$. The metallicity is $Z=0.1$, the
  number of grid points has been set to $N=250$. Note that the
  lightcurve resembles the instabilities present in Fig.~\ref{fig:scurve3}.} 
  \label{fig:pspectra6}
\end{figure}

A sample lightcurve and power spectrum for a 10 M$_\odot$ black hole is
shown in Fig.~\ref{fig:pspectra}. We binned the FFT data in
logarithmically equispaced bins of width $\Delta \nu/\nu=0.005$. The
disc undergoes limit cycles on a timescale of roughly 600 s. Note that
besides the fundamental frequency and their higher harmonics there are
more peaks present. These arise from the asymmetry of the lightcurve.

The luminosity of the disc is given in terms of $\dot
M_\textrm{obs}/\dot M$, where $\dot M$ is the external/average
accretion rate and $\dot
M_\textrm{obs}$ the ''observed'' accretion rate. This ratio is
equivalent to the ratio between the actual luminosity and the
luminosity of a stationary disc where no
advection is taken into account. 

\section{Comparison with previous work} 
\label{sect:review}

In this Section we briefly compare our models with previous and recent
work in the field. We restrict ourselves to optically thick disc
models.

There is no absolutely correct way of doing 1D accretion disc models
in the one-zone approximation, and in this paper we have adopted the
simplest approach, by ignoring all the details of the vertical
structure. To do it properly clearly requires doing two-dimensional
hydrodynamics. In Appendix~\ref{sect:appenergy} we derive the LHS of
the energy equation~(\ref{eq:energy}) explicitly from first principles.
We work in terms of the total internal energy and the mass of a disc
annulus. All other variables can be expressed in terms of mass and
internal energy of a disc annulus.

In early work on black hole accretion discs,
\citet{1991PASJ...43..147H} presented disc models with an energy
equation containing the evolution of the total energy. They apply
correction factors to try to take account of details of the vertical
disc structure. They mainly consider $M = 10 M_\odot$ and $\alpha=0.1$
but make use of a modified ''$\alpha$-$P$''-description by multiplying
the $r\phi$-component of the viscous stress-tensor with a factor
$\beta_P^q$. (Here $\beta_P$ is the ratio of gas pressure related to
total pressure, so that $q=0$ corresponds to what we use here, and
$q=1$ corresponds to the $r\phi$-component being proportional to
$\alpha P_{\rm gas}$.) For $q\geq 0.5$ they find stable disc solutions
while for $q<0.5$ they find limit-cycle solutions of increasing
amplitude. Their $q=0$ results are compatible with ours.

In a series of papers
\citet{1997MNRAS.287..165S,1998MNRAS.298..888S,2001MNRAS.328...36S}
take a similar approach to \citet{1991PASJ...43..147H} except that
they use different numerical factors to try to take account of the
details of the vertical disc structure. For this reason, while their
results are in line with those of \citet{1991PASJ...43..147H} and
those presented here, they differ in some minor details. In
particular, they find unstable solutions for a 10 M$_\odot$ black hole
and $\alpha=10^{-3}$ for $0.09<L_{\rm Edd} <1$ which is in line with
our models except the high-$\dot M$ end, where we find stability at 1
$L_{\rm Edd}$. They subsequently develop a scheme where they take
account of the possibility that the disc might become optically thin
and there is non-Keplerian rotation. 

\citet{2000ApJ...535..798N} present a model with the specific
application to GRS1915+105. They additionally include thermal
and radiation diffusion in radial direction. For the simulations
presented in this paper, however, these terms never become
important. They might become comparable to the terms containing the
radial advection of the internal energy, but this only will result in
factors of two at most. In the context of working in the
one-zone-approximation, the omission of these terms in our simulations
seems to be justified. They investigate models with different radially
varying values for the viscosity parameter $\alpha$. They furthermore try to
account for additional cooling in a corona and include some
flickering. Their flickering is a random ad-hoc modulation of the efficiency
of radiation coming from the inner parts of the disk. Lightcurves
similar to GRS1915+105 can be produced. No further
diagnostics of the flickering process (i.e. power density spectrum) are
tested there.   

\citet{2002ApJ...576..908J} consider a similar approach to the one
presented in this paper. They write the change in specific heat in
terms of temperature and density (i.e. surface density and scale
height, setting the radial advection of the scale height to
zero). They apply yet another set of correction factors to try
to take account of the vertical structure. 
Their results are in line with
those obtained by \citet{1997MNRAS.287..165S,1998MNRAS.298..888S,2001MNRAS.328...36S}.

In conclusion, these various approaches differ mainly in the nature
and actual numerical values of the correction factors (being in the
range 0.2 -- 16) applied when trying to take some account of the
vertical disc structure. There seems to be no unanimity on the values
that should be applied. In this paper we have taken the simple
approach of setting all these factors to unity. 

Some of the work reviewed here
\citep{2000ApJ...535..798N,2002ApJ...576..908J} also include prescriptions for energy loss to a wind. While there is
agreement that mass loss does not influence the results, energy loss
(cooling) does influence the disk and stabilises it. Energy loss
usually is only parametrised with a fraction of the radiative losses
taken away to the corona or the wind. This fraction is either set
constant or depends on the accretion rate \citep[e.g.][]{2002ApJ...576..908J}.
Without the energy loss to a wind, our results so far are in agreement
with the results in literature. 

In the next section we shall introduce our model for the flickering
which contains a self-consistent coupling between the mass, energy and
angular momentum loss of the disc to a wind coupled to the physical model
for the flickering.

\section{The Model for the flickering} 
\label{sect:flickering}


\subsection{The basic mechanism}

We now address the means of introducing a physical mechanism to produce
the observed short-timescale variability in black hole accretion
discs. As we mentioned in the Introduction the main problem is to find
a mechanism which gives the right amplitude and the right timescale.

\citet{1997MNRAS.292..679L} showed that if the viscosity parameter
$\alpha$ changes in a spatially uncorrelated manner on the local
viscous timescale at each radius, then the characteristic flickering
spectrum and the long timescales can be produced, but was unable to
suggest a physical mechanism which would produce this result.
\citet{2004MNRAS.348..111K}, following \citet{2003ApJ...593..184L},
put forward a solution. The main problem is that local variations in
$\alpha$ are correlated with small variations in the disc's internal
magnetic field which is generated by a local disc dynamo. This
operates characteristically on about the local dynamical
timescale. They envisage the disc dynamo operating as a set of
essentially independent dynamo cells with characteristic radial width
comparable to the local disc thickness. They then propose that the
dynamo mechanism also gives rise to a small poloidal field in each
cell. Most of the time this local poloidal flux is randomly aligned
from cell to cell and so has no net effect.  But from time to time
this poloidal field magnetic field is sufficiently aligned
from cell to cell that it gives rise to a large scale magnetic field
which is able to generate an outflow (wind or jet) which removes
angular momentum from the disc. Since the local field in each dynamo
cell changes on about the local dynamical timescale, this overall
alignment only happens on a large multiple of the dynamical timescale
and thus can be comparable or even longer than the viscous timescale.

We now apply these ideas to the thermal disc structure developed
above.  We stress here the additional concepts required in applying
this model to a realistic disc. More details of the underlying ideas
can be found in \citet{2004MNRAS.348..111K}. In the following subsections we
describe the details of this model and the equations involved.

\subsection{Evolution of the poloidal magnetic field}

We consider a poloidal magnetic field $B_z$ which evolves
according to the induction equation
\begin{equation}
  \label{eq:induction}
\frac{\partial B_z}{\partial t}=\frac{1}{R}\frac{\partial }{\partial
  R}\left(R\eta^*\frac{\partial B_z}{\partial
    R}\right)-\frac{1}{R}\frac{\partial}{\partial R}\left(R B_z U_R\right)+S_B.
\end{equation}
We allow the field to diffuse radially with an effective magnetic
diffusivity $\eta^*$ and allow for radial advection at velocity $U_R$
which is due to a magnetic wind torque. $S_B$ stands for a source term
for the poloidal field which is assumed to be generated by the local
magnetic dynamo. At the outer disc boundary we take $B_z=0$, allowing
magnetic flux to diffuse outwards, but preventing inward flux
advection.

\subsection{The source of poloidal field, $S_B$} \label{sect:sourceb}

The poloidal field $B_z$ is assumed to be generated by the disc dynamo
which generates the main disc viscosity. The magnetic field in the
disc, partitioned into disc annuli of radial extent $\Delta R\approx
H$, is assumed to change independently in each dynamo cell in a
stochastic fashion on the characteristic dynamo time scale
\begin{equation}
  \label{eq:dynamo}
  \tau_\textrm{d}=k_\textrm{d} \Omega_\textrm{K}^{-1},
\end{equation}
which we take to be a multiple of the local dynamical time scale
\citep[we take $k_\textrm{d}=10$,
cf.][]{1992MNRAS.259..604T,1996ApJ...463..656S}

Each annulus of the disc is assumed to generate a sufficiently large
internal tangled magnetic field of strength $B_\textrm{disc}$ which
generates the effective disc viscosity with $\alpha$ parameter
\citep{1973A&A....24..337S}
\begin{equation}
  \label{eq:alphadef}
  \alpha=\frac{B_\textrm{disc}^2}{4\pi P}.
\end{equation}
Although we assume that each dynamo cell generates a local poloidal
field $B_z$, we expect the magnitude of the poloidal component to be
small. We define a parameter $\beta_s$ so that the maximum value of
the poloidal field, $B_{z,max}$ is given by
\begin{equation}
  \label{eq:bzmax}
B_{z,max} = \beta_s B_{\rm disc},
\end{equation}
where $B_{\rm disc}$ is given by Equation~\ref{eq:alphadef}. We expect
$\beta_s \ll 1$. 

We then model the stochastic nature of the local poloidal field in the
following way. At each radius we define a set of times, $t_n = n
\tau_d, n = 1, 2, \ldots$ as integral multiples of the local dynamo
timescale at which the field changes. At each of these times we
generate a random number, $u(t_n)$, according the the Markoff process
\begin{equation}
  \label{eq:markoff}
  u(t_{n+1})=-\alpha_\textrm{1} u(t_n)+\epsilon(t_n)\;,
\end{equation}
where $\epsilon(t_n)$ is a random variable of zero mean and unit
variance, and we take $\alpha_\textrm{1}=0.5$.  The new value of the
magnetic field strength in the disc annulus then is given by
\begin{equation}
  \label{eq:bznew}
  B_{z}(t_{n+1})=u(t_{n+1})B_{z,\textrm{max}}.
\end{equation}

\begin{figure}
  \centering
  \includegraphics[angle=-90,width=0.5\textwidth]{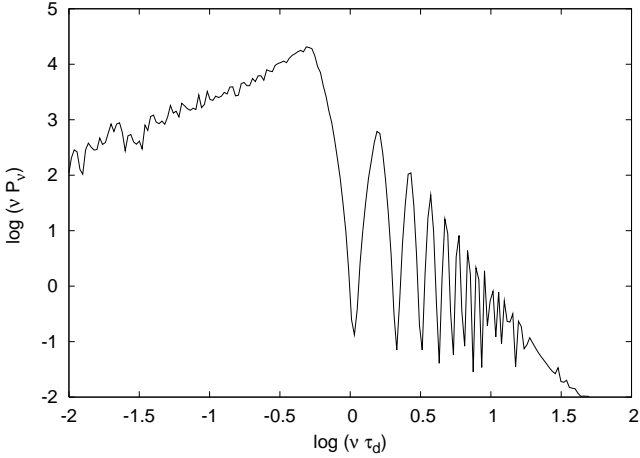}
  \caption{The power spectrum is shown for a sample time-series of a
  Markoff chain series (see eq.~\ref{eq:markoff}) for
  $\alpha_1=0.5$. This series is used to mimic the effects of locally
  acting magnetic dynamoes. In this figure we show the power spectrum
  for a dynamo acting on a timescale $\tau_d$. To account for
  timescales smaller than $\tau_d$, we linearly interpolated the
  Markoff chain. The frequency is normalised to $\tau_d$. The peak at
  the dimensionless frequency $\log (\nu \tau_d)=0$ is clearly visible
  including higher harmonics. Most of the power is released on
  slightly longer timescales (approximately $2.5 \tau_d$) coming from
  the broad peak at $\log (\nu \tau_d)\approx -0.4$.}
  \label{fig:markoff}
\end{figure}

We show a power spectrum of a sample time-series for $u(t_n)$ for a
magnetic dynamo acting locally on the dynamo time scale $\tau_d$ in
Fig.~\ref{fig:markoff}. For the purposes of the Figure, in order to
resolve shorter timescales than the dynamo timescale, we applied a
linear interpolation between the timesteps. We note that the power
spectrum peaks at a period of a few times $\tau_d$.

As mentioned above, in order to ensure the consistency of the model,
we need to make sure that the poloidal field $B_z$ is always smaller
than the disc magnetic field. This implies $\beta_\textrm{S}^2\ll 1$, so that
energetically the poloidal field is negligible. Thus this assumption
furthermore allows us to neglect the energy generation/loss due to our
dynamo process. \citet{1996MNRAS.281..219T} estimate this fraction to
be $\sim H/R$.

\subsection{Magnetic torque and the wind}

We assume that angular momentum loss occurs due to a magnetic wind/jet
when the poloidal field is of sufficiently large scale. This is likely
to occur when the poloidal fields generated by the individual dynamo
cells are, by chance, spatially correlated over a radial extent of the
order of $R$.  To measure the degree of spatial correlation in the
simulation, we define the radial averages
\begin{eqnarray}
  \left<B_z\right>&=&\frac{\int_{R-\Delta^-}^{R+\Delta^+}B_zR
    dR}{\int_{R-\Delta^-}^{R+\Delta^+}R dR},\\
  \left<B_z^2\right>&=&\frac{\int_{R-\Delta^-}^{R+\Delta^+}B_z^2R
    dR}{\int_{R-\Delta^-}^{R+\Delta^+}R dR}.
\end{eqnarray}

We choose the total interval over which we measure the correlation as
equal to $R$, that is we take
\begin{equation}
\Delta^-+\Delta^+=R.
\end{equation}
We then choose $R$ to be the geometric mean of $R - \Delta_-$ and $R +
\Delta_+$. Thus we choose the dimensionless constant $a$ so that
$R-\Delta^-=R/a$, $R+\Delta^+=aR$ with $a>1$. This then implies that
$a=(1+\sqrt{5})/2$, the golden ratio.

Thus we obtain
\begin{eqnarray}
  \left<B_z\right>&=&=\frac{\int_{R-\Delta^-}^{R+\Delta^+}B_zR
    dR}{R^2\left(a^2-1/a^2\right)},\\
    \left<B_z^2\right>&=&=\frac{\int_{R-\Delta^-}^{R+\Delta^+}B_z^2R
    dR}{R^2\left(a^2-1/a^2\right)}\;,
\end{eqnarray}
where we note that $a^2 - 1/a^2 = \sqrt{5}$.

We now define the quantity 
\begin{equation}
  \label{eq:coherence}
  Q=\frac{\left<B_z\right>^2}{\left<B_z^2\right>},
\end{equation}
which represents the degree of spatial correlation of the poloidal
magnetic field in neighbouring disc annuli. We note that $0 \leq Q \leq
1$, where $Q=1$ represents a fully coherent magnetic field, while for
$Q=0$ the magnetic field is fully randomly ordered.

The poloidal magnetic field $B_z$ produces a torque $T_\textrm{mag}$
on a disc annulus, width $\Delta R$, of size
\citep{1992MNRAS.259P..23L}
\begin{equation}
\label{eq:tmag} 
T_\textrm{mag}=-4\pi R^2
  \left(\frac{\left<B_z\right>^2}{4\pi}\right)\Delta R.
\end{equation}

Considering the change of angular momentum in the disc annulus, this
can be combined to give
\begin{equation}
2\pi R \Delta R \frac{\partial }{\partial t}\left(\Sigma R^2
  \Omega_\textrm{K}\right)+\Delta \left(2\pi R \Sigma v_R R^2
  \Omega_\textrm{K}\right) = \Delta G + T_\textrm{mag},
\end{equation}
where the terms on the LHS describe the local change of angular
momentum and the radial advection of angular momentum from neighbouring
annuli, while the terms on the RHS describe the viscous and magnetic
torques. Note that while angular momentum transport due to radial
advection and the viscous torque for the whole disc reduce to the
values at the inner and outer boundary, the magnetic torque produces a
local angular momentum loss term across the disc. 

Given this additional term ($T_\textrm{mag}$) in the angular momentum,
equation~(\ref{eq:angmom1}) now becomes
\begin{equation}
  \label{eq:angmom2}
  \frac{\partial }{\partial t}\left(\Sigma R^2
  \Omega_\textrm{K}\right)+\frac{1}{R}\frac{\partial }{\partial R}\left(R \Sigma v_R R^2
  \Omega_\textrm{K}\right)=\frac{1}{2\pi R}\frac{\partial G}{\partial
  R}-2 R \left(\frac{\left<B_z\right>^2}{4\pi}\right).
\end{equation}

In contrast to \citet{2004MNRAS.348..111K}, we now take explicit
account of the fact that the wind/jet, which removes angular momentum
from the disc, may also remove a significant amount of mass. We assume
that the magnetic outflow at radius $R$ co-rotates with the disc along each field
line, out to some Alfv\'{e}n radius $R_\textrm{A}$($>R$). Then the local mass loss rate,
$L_\Sigma$, is related to the torque by
\begin{equation}
  \label{eq:massloss}
  2\pi R \Delta R L_\Sigma R_\textrm{A}^2\Omega_\textrm{K}  =
  T_\textrm{mag}, 
\end{equation}
which implies that 
\begin{equation} L_\Sigma = - \frac{2
\left(R/R_\textrm{A}\right)^2}{\Omega_\textrm{K} R}
\left(\frac{\left<B_z\right>^2}{4\pi}\right).
\end{equation}

The loss term leads to a modification of the continuity equation
(\ref{eq:kg1}) in the form
\begin{equation}
  \label{eq:kg2}
  \frac{\partial \Sigma}{\partial t}+\frac{1}{R}\frac{\partial
  }{\partial R}\left(R\Sigma v_R\right)=L_\Sigma.
\end{equation}

Eqns. \ref{eq:kg2} and \ref{eq:angmom2} now can be combined using the
functional form of the viscous torque (eq.~\ref{eq:torque}) to give
\begin{equation}
  \label{eq:evol2}
  \begin{split}
  \frac{\partial \Sigma}{\partial t}&=\frac{3}{R}\frac{\partial
  }{\partial R}\left(\sqrt{R}\frac{\partial }{\partial R}\left(\nu_\textrm{t}
      \Sigma \sqrt{R}\right)\right)\\
&+\frac{4}{R}\frac{\partial}{\partial
    R}\left(\left(\frac{\left<B_z\right>^2}{4\pi}\right)
    \sqrt{\frac{R^5}{GM}}\left(1-\left(\frac{R}{R_\textrm{A}}\right)^2\right)\right)\\
  &-2\left(\frac{\left<B_z\right>^2}{4\pi}\right)\left(\frac{R}{R_\textrm{A}}\right)^2\sqrt{\frac{R}{GM}}.
  \end{split}
\end{equation}
For $R=R_\textrm{A}$ the enhancement of angular momentum due to the
extra torque $T_\textrm{mag}$ is balanced by the loss of angular
momentum taken away by the wind since it is taken away with the same
lever arm (see eq.~\ref{eq:massloss}) and the poloidal magnetic field
influences the evolution only through the mass loss. Note that
\citet{2004MNRAS.348..111K} essentially assumed that
$R_\textrm{A}=\infty$, and thus that mass loss is not important in the
model presented there. However we retain the ratio $R_\textrm{A}/R$
as a parameter and allow for mass loss, since analytical estimates
\citep{1986ApJ...301..571P} give $R_\textrm{A}/R=10$ while numerical
simulations \citep{1997ApJ...482..712O} suggest $R_\textrm{A}/R\approx
1.5$. Unless otherwise stated we choose $R_\textrm{A}/R=3$ as the fiducial
value.

The mass lost also removes the internal energy of the gas at a
rate
\begin{equation}
  \label{eq:windloss}
  Q_\textrm{w}=L_\Sigma U\approx L_\Sigma c_s^2,
\end{equation}
and thus we also require a revised energy equation (see
eq.~\ref{eq:energy}) which is in the form
\begin{equation}
  \label{eq:energy2} \begin{split} \frac{dq}{dt}&=\dot e +
  v_R\frac{\partial e}{\partial R} +2\pi R \Delta R P \dot H + P
  \frac{\partial \left(2\pi R H v_R\right)}{\partial R}\\ &=2\pi R
  \Delta R \left(Q^++Q_\textrm{w}-Q^-\right).
\end{split}
\end{equation}

We now have three equations (eqns.~\ref{eq:evol2}, \ref{eq:induction}
and \ref{eq:energy2}) describing the evolution of the disc. These are
integrated in time using a one-step Euler scheme. Advection is treated
in a first-order, upwind donor cell procedure. We note that the main
observational output from the disc is the radiation $L_\textrm{rad}$
coming from the disc (see eq.~\ref{eq:lrad}).

\subsection{Magnetic diffusivity}
Following \citet{2004MNRAS.348..111K}, we take the effective magnetic
diffusivity $\eta^*$ in eq.~\ref{eq:induction} to be
\begin{equation}
  \label{eq:diffusivity}
  \eta^*=\nu_\textrm{t} \max\left(1,QR/H\right),
\end{equation}
where we assume that the Prandtl number is of the order unity
($\eta=\nu_\textrm{t}$), but allow it to be enhanced by a factor $R/H$
\citep{1989admf.proc...99V,1994MNRAS.267..235L} and include large
scale effects by multiplying it by the coherency factor $Q$.  For
further discussion see \citet{2004MNRAS.348..111K}.

\subsection{The interplay between the timescales for the flickering}\label{sect:interplay}

In this section we quantify the timescale argument for the appearance
of flickering. As mentioned previously, fluctuations propagate through
the disc, if the viscous timescale is shorter than the magnetic
timescale, i.e, the timescale where we expect sufficient alignment of
the poloidal magnetic field in
neighbouring cells. 

\begin{figure}
  \begin{center}
  
    \includegraphics[angle=-90,width=0.49\textwidth]{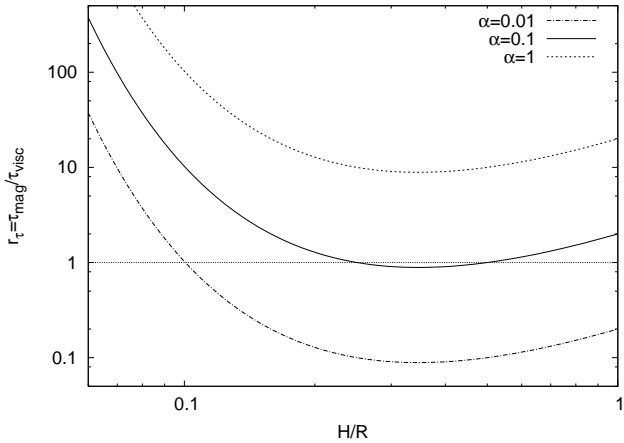} 
  \end{center}
  \caption{Ratio $r_\tau$ of the magnetic and viscous timescale in
    dependence of $H/R$ (see eq.~\ref{eq:propcond2}) for different
    values of the viscosity parameter $\alpha$ and $k_d=10$. Only for
    $r_\tau\geq 1$ fluctuations can propagate. }  \label{fig:rat-timescale}
\end{figure}

The viscous timescale is given by \citep{1981ARA+A..19..137P}
\begin{equation}
  \label{eq:tvisc}
  \tau_\textrm{visc}=\frac{1}{\alpha\left(H/R\right)^2}\tau_\textrm{dyn}\;,
\end{equation}
where $\tau_\textrm{dyn}=1/\Omega_K$ stands for the dynamical timescale. $H/R$ is
the opening angle of the disc. The
magnetic timescale, i.e. the timescale, on which we expect
neighbouring field lines to be sufficiently aligned, is given by
\citep{2003ApJ...593..184L}
\begin{equation}
  \label{eq:taumag}
  \tau_\textrm{mag}=2^{R/H}k_d \tau_\textrm{dyn}\;.
\end{equation}
Thus the condition for propagation is 
\begin{equation}
  \label{eq:propcond}
  r_\tau=\tau_\textrm{mag}/\tau_\textrm{visc}\geq 1\;, 
\end{equation}
which translates into
\begin{equation}
  \label{eq:propcond2}
  r_\tau=2^{R/H}k_d  \alpha\left(H/R\right)^2\geq 1\;.
\end{equation}
We plot the ratio $r_\tau$ in Fig.~\ref{fig:rat-timescale} for
different values of $\alpha$. The value of $k_d$ is fixed to 10 (see Sect.~\ref{sect:sourceb}). The ratio
is smallest for $(H/R)_\textrm{min}\approx 0.35$. For $\alpha=0.1$
fluctuations formally propagate only for $H/R\gtrsim 0.5$ or $H/R\lesssim
0.25$ only, while for $\alpha=1$ fluctuations can propagate for all
$H/R$. In contrast to that, for $\alpha=0.01$ fluctuations only
propagate for $H/R \lesssim 0.1$. 

This is in line with findings of
\citet{2001MNRAS.321..759C} who state that fluctuations close to the
dynamical timescale are
effectively damped in geometrically thin discs but can survive in
geometrically thick discs since then the fluctuation timescale becomes
comparable to the viscous timescale. In our model
fluctuations formally survive for lower $H/R$ ratios as well. Then,
however, the magnetic timescale is so long that alignment becomes extremely
rare. From the above analysis this probably puts a lower limit on the
viscosity parameter $\alpha$ of $\alpha \gtrsim 0.1$ since most of the
variability is expected to come from a optically thin, but
geometrically thick disc.

\section{Observations of the flickering in the thermal component of
the X-Ray spectrum}

\label{sect:obs}

Before describing the numerical simulations we first consider
the likely observational scenario to which they are relevant.  In
generalising the work of \citet{2004MNRAS.348..111K} to more
physically realistic discs, we have initially confined ourselves to
considering the standard accretion disc configuration, with the
addition of heat advection. In general this is assumed to correspond
in X-ray binaries to the thermally dominant (TD) or high/soft
state. However, from an observational point of view, the variability
in the TD or high/soft state is much less than in the LH state.

Usually the main components in the spectrum of X-Ray binaries can be
well fit by a black-body and a power-law component
(cf. \cite{0306213}). The black-body or thermal component for galactic
X-Ray binaries appears at energies around 1 keV, while the power-law
component extends to higher energies. The components are thought to
represent the radiation from the optically thick disc and a optically
thin corona, respectively.
Thus, since in the current paper, we only attempt to describe the
time-dependent behaviour of the thermal disc component, i.e. the
optically thick disc, we need to restrict ourselves to observations
which explicitly concentrate on measuring the flickering in the
thermal disc component of the X-Ray spectrum.

\citet{2001MNRAS.321..759C} show a power spectrum density (PSD) of Cyg
X-1 in the high-soft state for energies of 6-13 keV, and state, that
the amplitude in the softer bands is much smaller due to the influence
of the black-body emission from the disc. They argue that their data
is consistent with essentially all of the variability being associated
with a power-law component and not with black-body emission (i.e. disc
component), but do not quote a formal limit.

\citet{1994ApJ...435..398M} examine Ginga observations of Nova Mus
1991 (GS 1124-683). They disentangle the variability in the disc and
power-law component by calculating the PSD for different observations
which have different fractional contributions from the power-law
component to the total flux. For a power-law component fraction above
approximately 10 per cent, the PSD at 0.3 Hz is an increasing function
of power law fraction. While for lower values it is roughly constant
(1.5$\cdot$10$^{-5}$ Hz$^{-1}$). At these low values, the slope of the
PDS is about -0.7 but due to the error bars it still could be fit with
-0.5. \citet{1999ApJ...510..874N} show that for counting noise
dominated time-series the expected PDS slope is $-0.5$. Thus we
conclude the data are consistent with a non-detection of flickering in
the disc component. By using the value of the PDS at 0.3 Hz only we
estimate an upper limit to the rms variability of $r<0.2$ per cent.

\citet{2001MNRAS.320..316N} consider LMC X-1 and LMC X-3. These are
persistent black hole sources (i.e. not transients), and generally
show spectra dominated by a soft, thermal component. They detect no
variability in LMC X-3 down to a level of about 0.8 per cent. 
They do detect variability in LMC X-1 of 7 per cent throughout the
spectral range considered (0-9 keV), but since for  LMC X-1 the power
law component is much stronger that in LMC X-3, this could just be
variability of the power law component.

Observations of the transient 4U 1543-47 during its 2002 outburst are
presented in \citep{2004ApJ...610..378P}. During most of the decay the
spectrum is soft and dominated by black body emission. The PDS shape
in this phase is roughly 1/$\nu$ and the variability is about $r=1$ per
cent. However the fraction of black body emission never exceeds 80 per
cent, i.e. the power law always contributes more than 20 per
cent. Comparison with the results of \citet{1994ApJ...435..398M}
shows, that for a black-body flux fraction of 20 per cent, if the relation
shown there is representative, the fractional rms must be $r=1$
per cent.

\citet{2004ApJ...603..231K} report on the light curves of a number of
transients observed with RXTE. They again find strong variability in
the low/hard state, while in the high/soft state they only can give
upper limits, since they are counting noise dominated. Depending on
the source, and the quality of the data, these limits are in the range
1 to 8 percent. 

\begin{figure} \centering
  \includegraphics[angle=-90,width=0.5\textwidth]{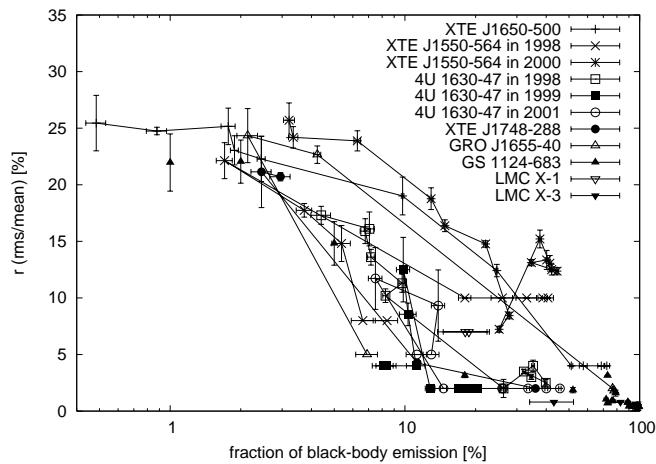}
  \caption{Fractional rms variability $r$ for different black-body
  flux fractions in the 3-25 keV band for different objects. All data
  from \citet{2004ApJ...603..231K} except the data for GS 1124-683,
  which have been taken from \citet{1994ApJ...435..398M} and LMC X-1
  and X-3 from \citet{2001MNRAS.320..316N}.}  \label{fig:rms-bb}
\end{figure}

To quantify the relation between the flux fraction and the rms
variability, we have taken data from \citet{2004ApJ...603..231K} and
\citet{1994ApJ...435..398M} and plot the rms versus the black body
flux fraction in Fig.~\ref{fig:rms-bb}. The fluxes are calculated in
the 3-25 keV band. Since \citet{1994ApJ...435..398M} only give the
normalised power density $P(\nu)$ at frequency $\nu = 0.3$ Hz, we
estimate the integrated rms r by using $r=2\sqrt{\nu P_{\nu}}$. The
factor 2 is introduced to account for the $d\nu/\nu$ in the definition
of the integrated rms (cf. eq.~\ref{eq:frms}). The black-body fraction
for the LMC X-1 and X-3 data of \citet{2001MNRAS.320..316N} has been
calculated using their black-body + powerlaw fits. While there is some
scatter in the data, the global trend is that for black-body flux
fractions lower than 5 per cent the rms variability is about 25 per
cent, while it drops beyond detectable limits for larger values of the
black body fraction. The transition is not very sharp, but occurs
around a black body fraction in the range 3 to 20 per cent. The
transition might be sharper than this but we are only using data in
the 3-25 keV range where the black-body contribution is small compared
to the power-law component, and varies from source to source.  In
addition, we note that \citet{2005MNRAS.360..825Z} carry out an
energy-dependent analysis of the fractional rms for GRS 1915+105 and
come to similar conclusions.

To summarise, it is clear from the observations that the rms
variability of the black-body component in the X-Ray spectra of black
hole sources is small. Indeed there are no clear detections, but only
upper limits. As a representative value we take $r=1$ per cent as an
upper limit for the variability of the thermal component.

\section{Results} 
\label{sect:res}

To demonstrate the effect of the stochastic magnetic wind/jet on our
disc models model with the physical input so far, we ran some models
to assess the influence of the parameters on the variability. 
For all models, we fix the mass of the black hole to be 10 $M_\odot$,
the viscosity parameter $\alpha=0.1$ \citep[appropriate for standard
optically thick discs, see][]{1997xrb..book.....L} and take for the dynamo
timescale (cf. eq.~\ref{eq:dynamo}) $k_d=10$.

\begin{figure*}
  \centering
  
    \includegraphics[angle=-90,width=\textwidth]{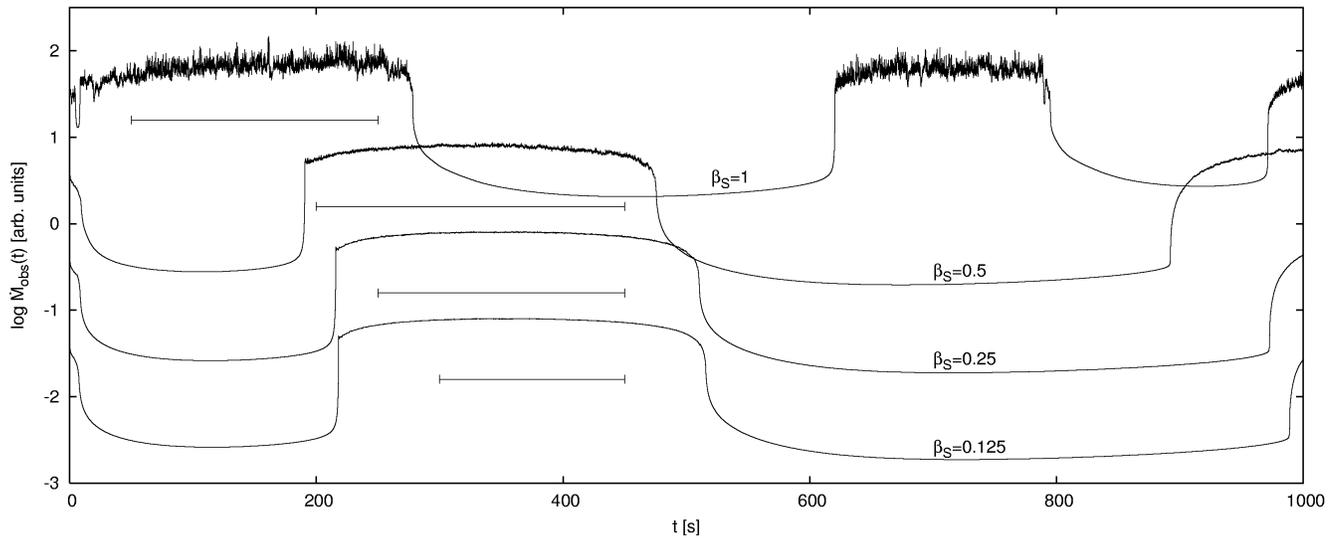}
  
  \caption{Sample lightcurves for a 10 M$_\odot$ mass black hole
    accreting at 0.5 $\dot M_\textrm{Edd}$ for different values of
    $\beta_\textrm{S}$, the strength of the poloidal field compared to the disc
    magnetic field. The ratio of the Alfv\'{e}n
    radius $R_\textrm{A}$ to the radial distance $R$ is $R_\textrm{A}/R=3$, the viscosity parameter $\alpha=0.1$.
    The intervals indicate the time-segments
    shown in Fig.~\ref{fig:samplight-det} used to calculate the power
    spectra shown in Fig.~\ref{fig:rms-bs}. } \label{fig:samplight}
\end{figure*} 

\begin{figure}
  \centering
  {\large 
  \begin{tabular}{l}
    $\beta_\textrm{S}=1$\\
    \includegraphics[angle=-90,width=0.49\textwidth]{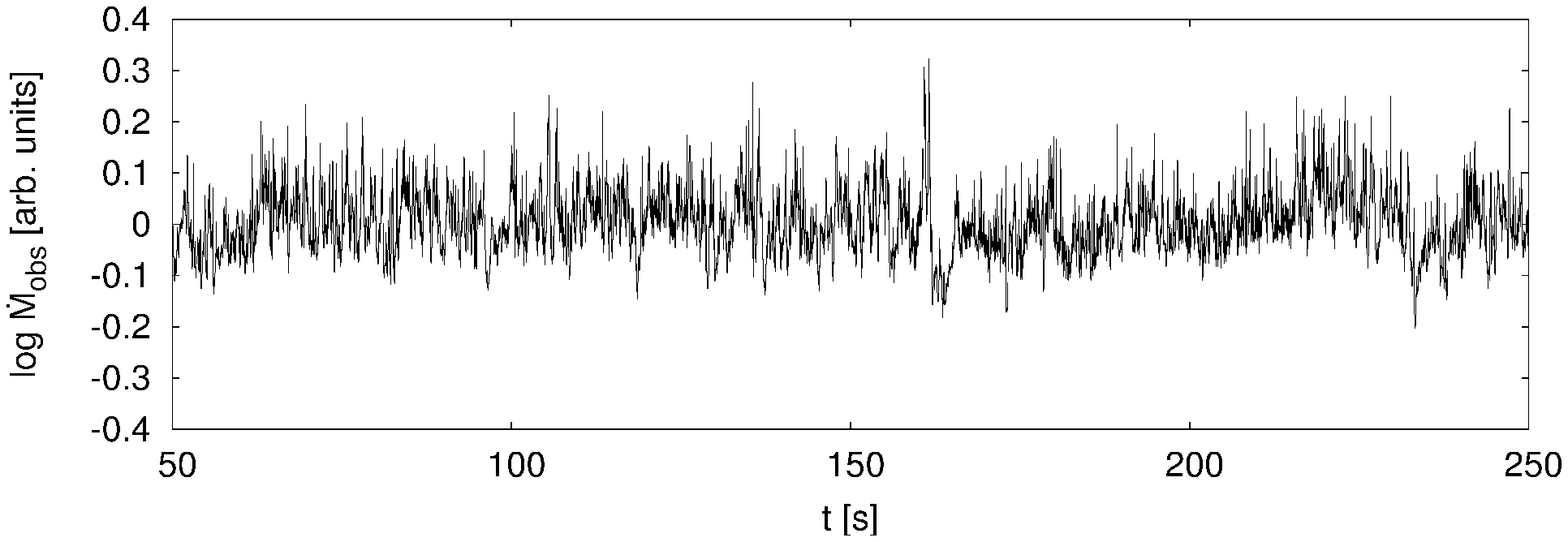}\\
    $\beta_\textrm{S}=0.5$\\
    \includegraphics[angle=-90,width=0.49\textwidth]{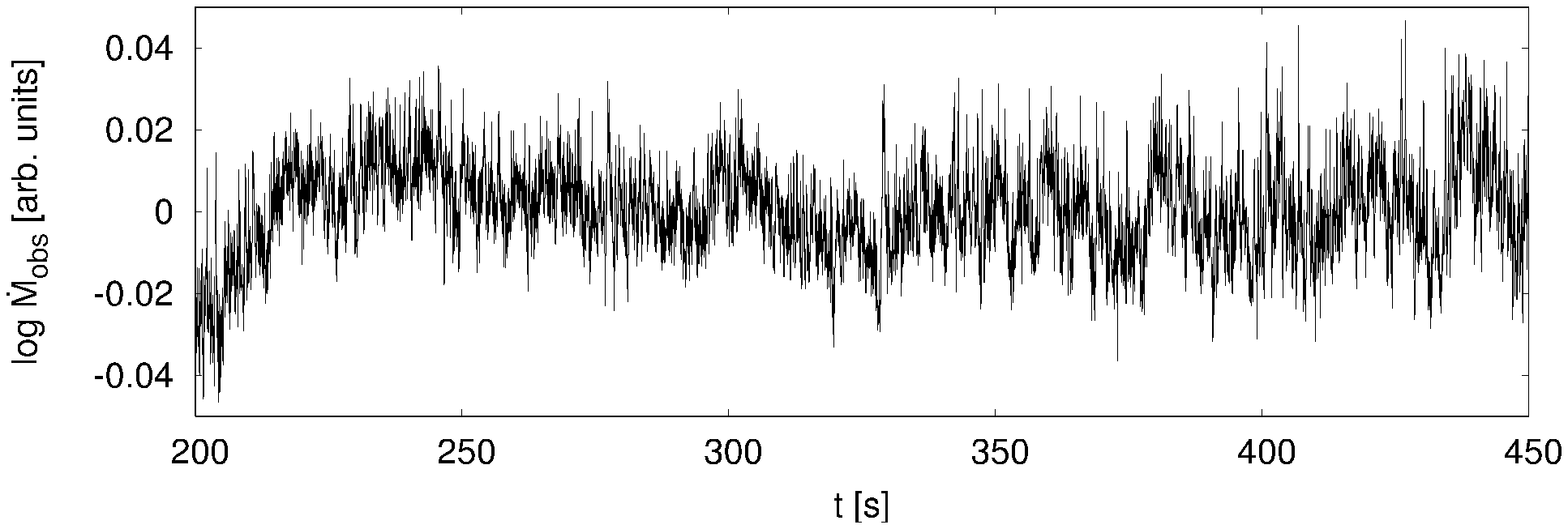}\\
    $\beta_\textrm{S}=0.25$\\
    \includegraphics[angle=-90,width=0.49\textwidth]{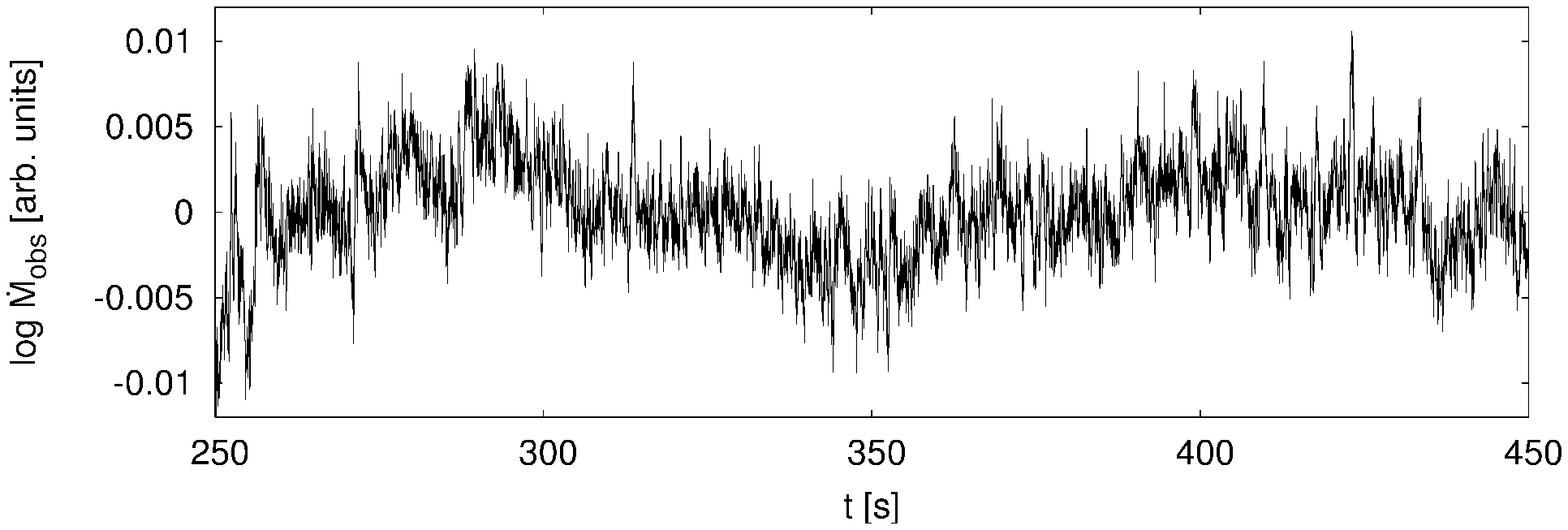}\\
    $\beta_\textrm{S}=0.125$\\
    \includegraphics[angle=-90,width=0.49\textwidth]{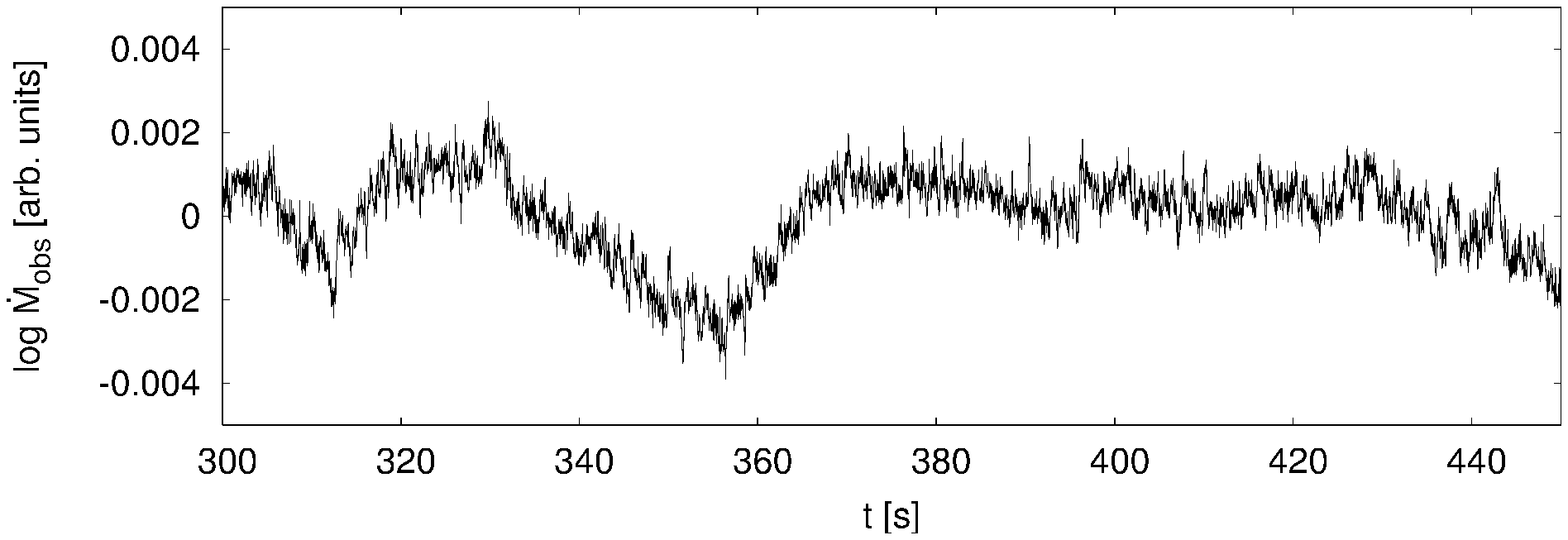}\\
  \end{tabular}}
  \caption{Lightcurve segments for the cases shown in
    Fig.~\ref{fig:samplight} for different $\beta_\textrm{S}$. These are
    used to calculate the PDS in Fig.~\ref{fig:rms-bs}. $\dot
    M_\textrm{obs}$ is given in arbitrary units, but the amplitude
    still has got some meaning. Note that the
    overall shape of the lightcurve does not change considerably,
    while the range in $\dot M_\textrm{obs}$ changes
    significantly. The longterm trend in the data has been removed.} \label{fig:samplight-det}
\end{figure} 

\begin{figure}
  \begin{center}
  
    \includegraphics[angle=-90,width=0.49\textwidth]{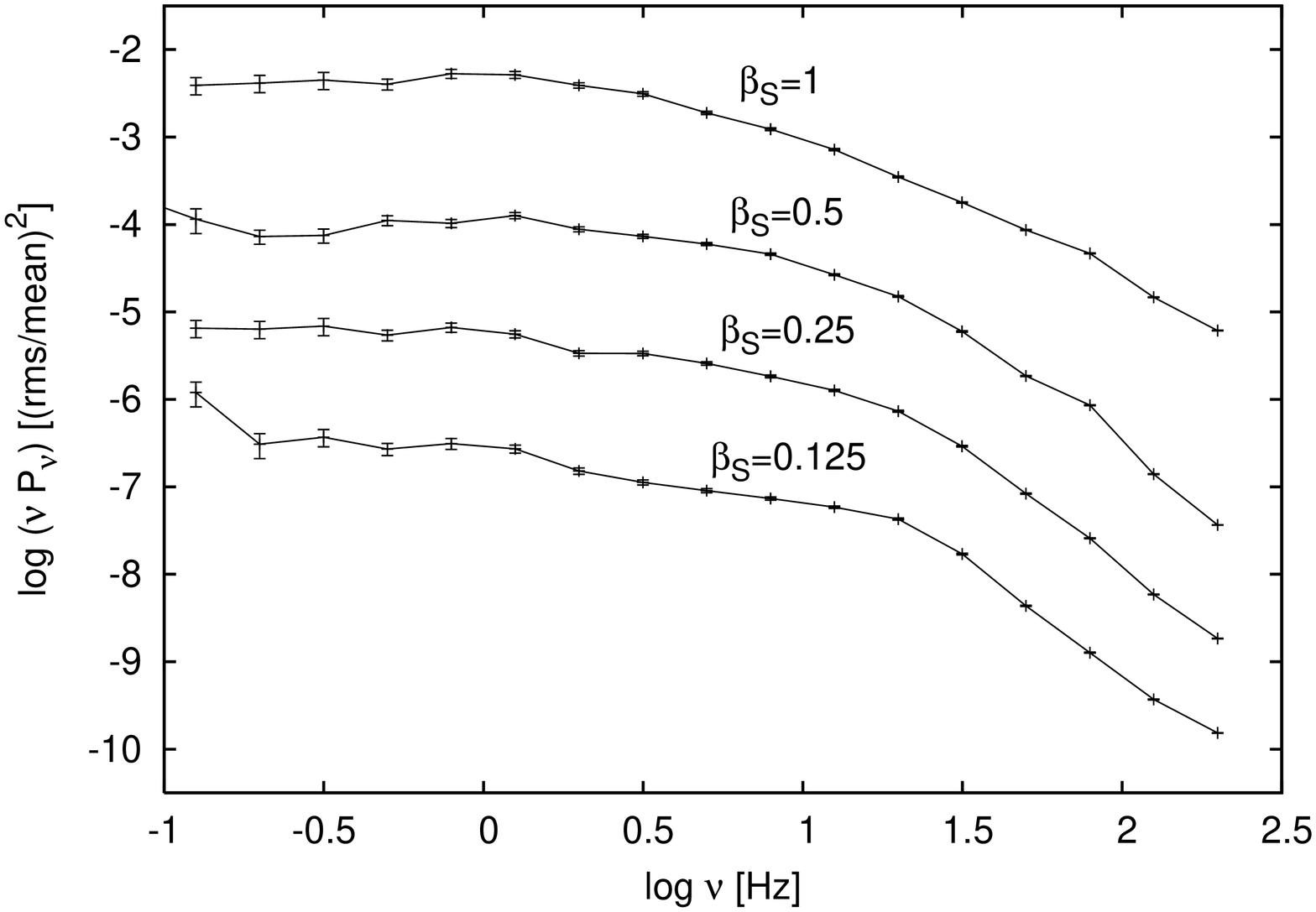} 
  \end{center}
  \caption{Power spectra for a 10 M$_\odot$ model solar mass
    black hole, accreting at 0.5 $\dot M_\textrm{Edd}$ for different
    values of $\beta_\textrm{S}$ for the lightcurve segments shown in Fig.~\ref{fig:samplight-det}. $\alpha=0.1$}  \label{fig:rms-bs}
\end{figure}

First we consider a black
hole accreting at half of the Eddington rate, $\dot{M} = 0.5
\dot{M}_{\rm Edd}$. We take the Alfv\'{e}n radius at each
radius to be a constant value of $R_\textrm{A}/R = 3$. From Figure 4, in the 
absence of magnetic flickering (i.e. with
$\beta_\textrm{S} = 0$) the disc is unstable and shows limit-cycle
behaviour. Fig.~\ref{fig:samplight} shows sample lightcurves for
different values of $\beta_\textrm{S}$. The limit cycle is still present,
i.e. the disc oscillates between a high $\dot M_\textrm{obs}$ and low
$\dot M_\textrm{obs}$ state. It is best evident from the $\beta_\textrm{S}=1$
case that while in the high $\dot M_\textrm{obs}$ state there is some
variability, in the low $\dot M_\textrm{obs}$ state there is hardly any sign
of variability. This is a result of the smaller $H/R$ ratio in the
inner disc during the low $\dot M_\textrm{obs}$ state. Then the
alignment timescale \citep[proportional to the factor $2^{R/H}$ times
the dynamical timescale, see][]{2003ApJ...593..184L} is much longer
than the viscous timescale (proportional to $(R/H)^2$ times the
dynamical timescale): Fluctuations formally could survive, but since
alignment occurs on extremely long timescales only, they are unlikely to
produce flickering \citep[see Section~\ref{sect:interplay} and][]{2004MNRAS.348..111K}. 
In the high-$\dot M$ state, however, both timescales are at least comparable
and so fluctuations propagate to the inner disc without being smeared out
by the viscosity.

\begin{figure}
  \begin{center}
    \includegraphics[angle=-90,width=0.49\textwidth]{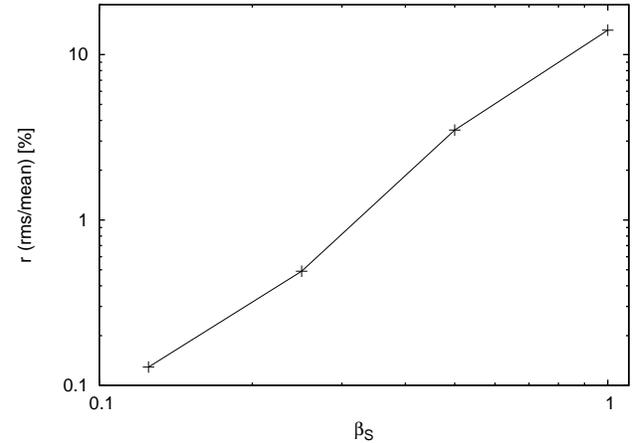}
  \end{center}
  \caption{Integrated fractional rms $r$ for a 10 M$_\odot$ model solar mass
    black hole, accreting at 0.5 $\dot M_\textrm{Edd}$ in dependence
    of $\beta_\textrm{S}$, the strength of the poloidal field compared to the
    disc magnetic field (cf. Fig.~\ref{fig:rms-bs}). The viscosity
    parameter is $\alpha=0.1$.}  \label{fig:rms-bs2}
\end{figure}

It is evident from Fig.~\ref{fig:samplight}, that a
non-zero $\beta_\textrm{S}$ influences the strength of the variability. In Fig.~\ref{fig:rms-bs} we show the influence of the parameter
$\beta_\textrm{S}$ on the power density spectra (PDS), while the integrated
fractional rms $r$ (see eq.~\ref{eq:frms}) is shown in Fig.~\ref{fig:rms-bs2}. 

The time segments used for
the calculation of the PDS are indicated in
Fig.~\ref{fig:samplight}. Throughout this section, we only take
segments from the lightcurve, where the disc was in the high-$\dot M$
state. Since for red-noise the lightcurve
is only a stochastic realisation of the underlying process, the
calculated PDS fluctuates randomly around the ''true''
spectrum \citep[e.g.][]{2002MNRAS.332..231U}. To overcome
this difficulty, we divided each segment in 10 equally spaced
subsegments, calculated the PDS of each subsegments and took the
average of these. We further normalised each subsegment to remove
linear trends in the lightcurve.

We see from Fig.~\ref{fig:rms-bs2} that the rms
variability is $r<1$ \%, for $\beta_\textrm{S}<0.2$ and that $r$ strongly
increases for larger $\beta_\textrm{S}$ and reaching 10-20 \% for
$\beta_\textrm{S}=1$. The near quadratic increase of $r$ with $\beta_\textrm{S}$
reflects the fact, that the poloidal magnetic field influences the
disc only through the torque which is proportional to
$\left<B_z^2\right>$ which in turn is proportional to $\beta_\textrm{S}^2 P$,
i.e. a fraction $\beta_\textrm{S}^2$ of the viscous torque (see
eqns.~\ref{eq:tmag}, \ref{eq:bzmax}, \ref{eq:alphadef} and \ref{eq:torque2}).

Given the
observational constraints, discussed in Section 7, it is evident that
we need to take $\beta_\textrm{S}$ to be smaller than around 0.1 - 0.2. Thus in
this model we require the energy density in the poloidal field
component to be at most only a few percent of the energy density in
the magnetic field generated by MRI in the disc.

\begin{figure}
  \begin{center}
 
    \includegraphics[angle=-90,width=0.49\textwidth]{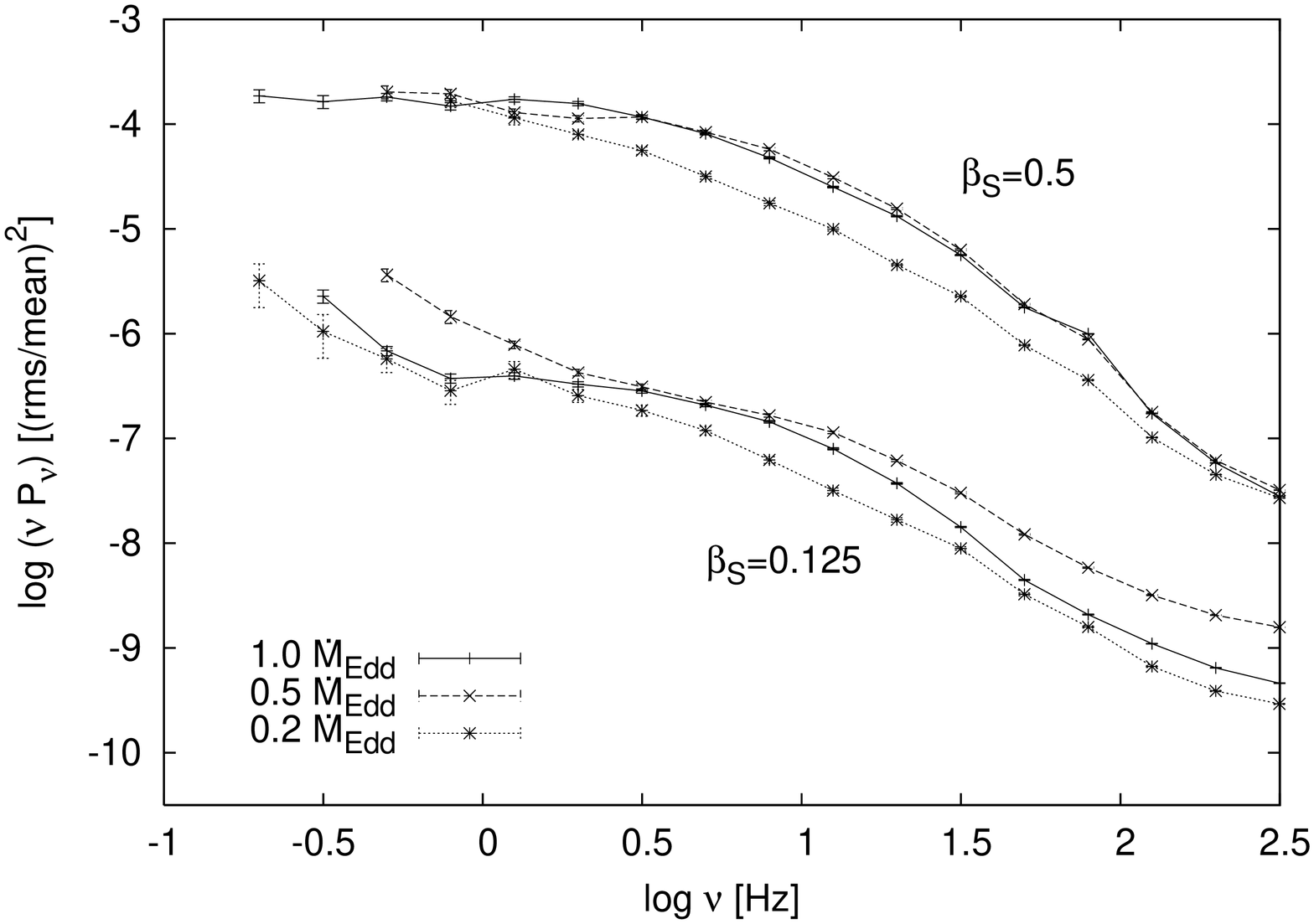} 
 
  \end{center}
  \caption{Power spectra for a 10 M$_\odot$ model solar mass
    black hole, accreting at different accretion rates for different
    $\beta_\textrm{S}=0.125$ and $0.5$. $\alpha=0.1$.}  \label{fig:rms-dm}
\end{figure}

Next we consider the influence of the accretion rate on the
variability for constant $\beta_\textrm{S}$. Fig.~\ref{fig:rms-dm} shows the
PDS for two values of $\beta_\textrm{S}$ for different accretion rates. While
the integrated rms variability $r$ does not depend on the accretion
rate but on $\alpha$, the shape of the PDS for different accretion
rates is slightly different.

\begin{figure}
  \begin{center}
    \includegraphics[angle=-90,width=0.49\textwidth]{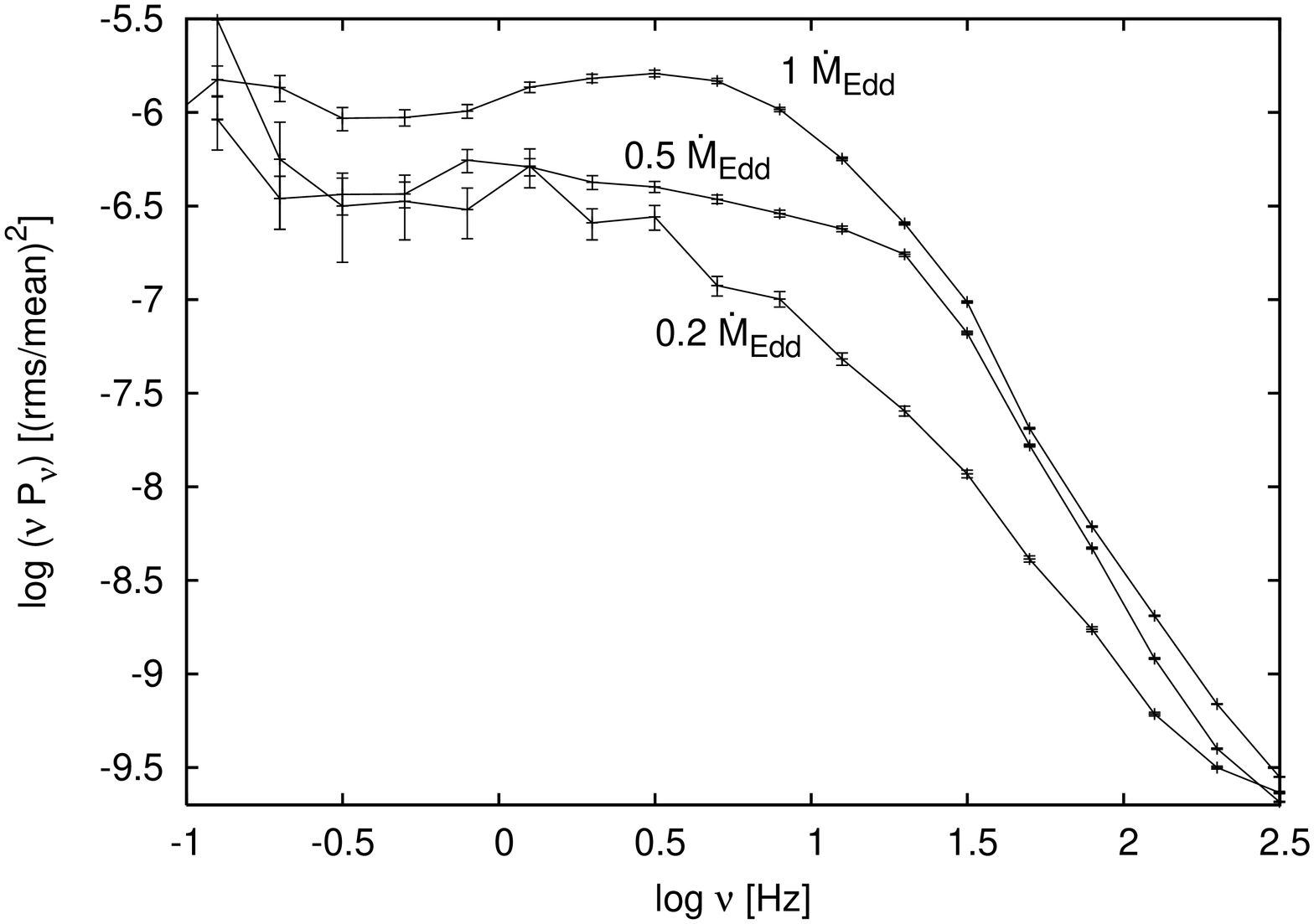} 
  \end{center}
  \caption{Power spectra for a 10 M$_\odot$ model solar mass
    black hole, accreting  at different rates for $\beta_\textrm{S}=0.5
    \left(H/R\right)$. $\alpha=0.1$.}  \label{fig:rms-bs-hr-0.5}
\end{figure}

\begin{figure}
  \begin{center}
    \includegraphics[angle=-90,width=0.49\textwidth]{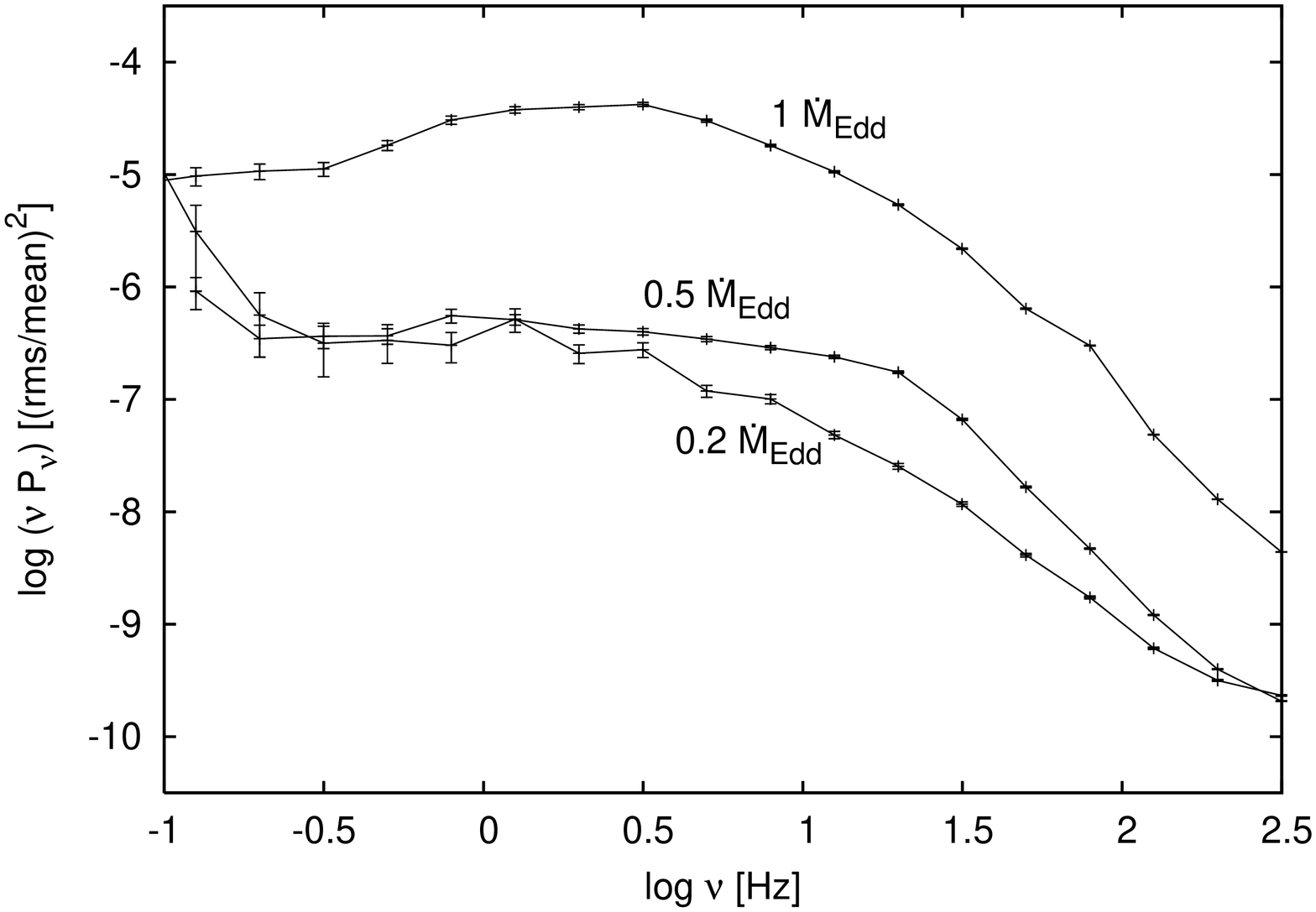}
  \end{center}
  \caption{Power spectra for a 10 M$_\odot$ model solar mass
    black hole, accreting  at different rates for $\beta_\textrm{S}=
    \left(H/R\right)$. $\alpha=0.1$.}  \label{fig:rms-bs-hr-1}
\end{figure}

We also briefly investigate the behaviour which occurs if $\beta_\textrm{S}$,
the ratio of poloidal and tangled magnetic field strengths, is assumed
to vary with $H/R$. For example, in their simple dynamo model,
\citet{1996MNRAS.281..219T} suggest that $\beta_\textrm{S}$ might be
proportional to $H/R$.  If so, since $H/R$ in the simulations
described is at most $H/R \sim 1/5$, we can put a constraint on the
constant of proportionality to be smaller than unity.

In Figures~\ref{fig:rms-bs-hr-0.5} and~\ref{fig:rms-bs-hr-1} we show 
the results of simulations with the
same parameters as before but with $\beta_\textrm{S} = 0.5 H/R$ and $\beta_\textrm{S} =
H/R$ for varying accretion rate. For the 1 $\dot M_\textrm{Edd}$ cases there
 is a significant decline of the
PDS for small frequencies. This decline is a result of a declining
$H/R$ in the outer disc which shortens the amplitude of flares
released from further out (i.e. longer timescales) through the
$H/R$-dependence of $\beta_\textrm{S}$. For the lower accretion rates this
decline is far less obvious, but then the time the disc is spending in
the high-$\dot M$ state is becoming shorter and so does the time
available for the FFT and averaging. The integrated fractional rms in
for the same accretion rate changes by a factor of about 4 when
comparing the cases of  $\beta_\textrm{S} = 0.5 H/R$ and $\beta_\textrm{S} =
H/R$ consistent with the result of Fig.~\ref{fig:rms-bs2} where we
show the quadratic dependence of $r$ with $\beta_\textrm{S}$.

\begin{figure*}
  \centering
  \begin{tabular}{cc}
    \includegraphics[angle=-90,width=0.5\textwidth]{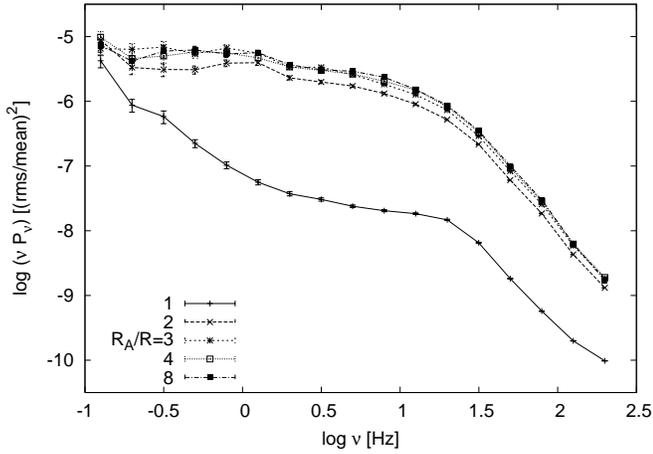} & \includegraphics[angle=-90,width=0.5\textwidth]{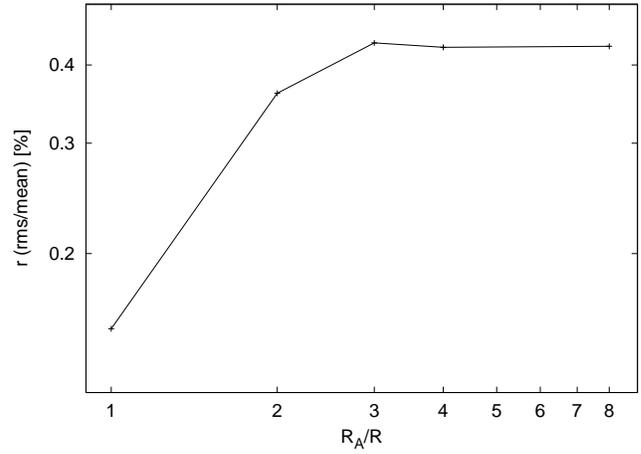}
  \end{tabular}

  \caption{(left)
    Power spectra for different values of $R_\textrm{A}/R$, the ratio of the
    Alfv\'{e}n radius $R_\textrm{A}$ to the
    radial distance $R$ for a
    10 $M_\odot$ black hole 
    accreting at 0.5 $\dot M_\textrm{Edd}$ with
    $\beta_\textrm{S}=0.25$, the ratio of the poloidal field compared to the
    disc magnetic field. (right) Integrated rms-variability $r$ as a function of $R_\textrm{A}/R$. For
    all $R_\textrm{A}/R> 3$ the power spectra and the value of $r$ is essentially
    indistinguishable.} \label{fig:rra}
\end{figure*} 

Finally we explored the influence of different values of
$R_\textrm{A}/R$ on the variability. In Figure~\ref{fig:rra} we show the
influence for a 10 $M_\odot$ black hole accreting at 0.5 $\dot
M_\textrm{Edd}$ with $\beta_\textrm{S}=0.25$. The system changes with $R_\textrm{A}/R$
increasing from a mass loss to a magnetic torque dominated regime. 
For the parameters chosen, 
the integrated rms-variability $r$ is increasing with $R_\textrm{A}/R$ and
saturates for $R_\textrm{A}/R>3$. The power spectral shape then is
indistinguishable within the error bars. Thus the assumption of
\citet{2004MNRAS.348..111K} who set $R_\textrm{A}/R=\infty$  is
justified as long as the ''real'' $R_\textrm{A}/R>3$. 

\section{Conclusions} \label{sect:conclusions}

We have generalised the model of \citet{2004MNRAS.348..111K} for variability in
black hole accretion discs by including proper consideration of the
local disc structure. We take the disc properties to depend only on
radius $R$, and so, in common with other authors, make a local
one-zone approximation for the disc structure.

In the absence of the addition of a stochastic magnetic dynamo, which
drives the flickering, we mainly reproduce results similar theoretical
work has shown.  We agree that
taking account to advective heat flow is of crucial importance in
determining the structure of the inner regions, but find that the
assumption of strictly Keplerian angular velocities, and the neglect
of radial pressure gradients, are generally reasonable
assumptions. 

When a stochastic dynamo is included we find behaviour similar to
that described by \citet{2004MNRAS.348..111K}. However, because the degree of
variability of the thermal disc is observed to be small (less than
around 1 per cent), if it is detected at all, we can only draw limited
conclusions. Our main finding is that for consistency we require that
the strength of the poloidal field generated by any radial disc dynamo
cell must be at least an order of magnitude smaller than the field
generated by the dynamo within the disc. The simulations reported here
however are of help to understand the influence of the parameters
on the variability.  

In this paper we have restricted ourselves to modelling just the
thermal disc component, and so have had to restrict our attention to
the high/soft or thermally dominant state of black hole binaries. In
future work we shall begin to include consideration of a hot corona,
in addition to the cooler thermal disc, so that we can begin to model
the more highly variable low/soft state of the black hole binaries, as
well as the X-rays from AGN \citep{2005MNRAS.359..345U}.

\section*{Acknowledgements}

MM gratefully acknowledges financial support from PPARC. We also
thank
the anonymous referee for comments on the paper which helped to
improve it significantly.

\appendix
\section{The energy equation} \label{sect:appenergy}

\subsection{Derivation of the LHS}

The first law of thermodynamics is 
\begin{equation}
dQ=dE+Pd\left(\frac{1}{\rho}\right),
\end{equation}
where $dQ$ stands for the change in heat $Q$ per unit mass, $dE$ the
change in internal energy $E$ per unit mass, and
$Pd(1/\rho)$ the volume work.
$P$ is the pressure and the  specific volume is $1/\rho$.

The continuity equation in general form reads
\begin{equation}
d\left(\rho V\right)= \rho dV+ Vd\rho = 0,
\end{equation}
which is equivalent to
\begin{equation}
  \frac{dV}{V}=-\frac{d\rho}{\rho}\;.
\end{equation}

In total differentials using the continuity equation, the energy
equation can be shown to be equivalent to
\begin{equation}
d\left(\rho V Q\right) = d\left(\rho V E\right) + P dV.
\end{equation}
This relation implies that the change of total heat in a volume $V$ is
given by a change in the total internal energy and the expansion of
the volume.

In cylindrical coordinates, vertically averaged, we have
\begin{equation}
V=2\pi R H \Delta R
\end{equation}
Thus
\begin{equation}
d\left(\pi R \Sigma \Delta R Q\right) = d\left(\pi R \Sigma \Delta R
E\right) + 2\pi R \Delta R P dH.
\end{equation}
Replacing the specific heat $Q$ by the total heat content $q$ in a
semidisc annulus, 
\begin{equation}
  q=\pi R \Sigma \Delta R Q,
\end{equation}
and the specific internal energy $E$ by the total internal energy $e$
in a semidisc annulus,
\begin{equation}
  e=\pi R \Sigma \Delta R  E\;,
\end{equation}
we can write
\begin{equation}
dq=de+  P d\left( 2\pi R  H \Delta R\right).
\end{equation}
Time-dependently, the LHS of the energy equation is written
\begin{equation}
\frac{dq}{dt}=\frac{de}{dt}+ P \frac{d\left( 2\pi R  H \Delta R\right)}{dt}.
\end{equation}
Taking into account advection, with radial velocity $u_R$,  
\begin{equation}
\frac{dq}{dt}=\dot e + u_R\frac{\partial e}{\partial R} +2\pi R \Delta
R P \dot H + P \frac{\partial \left(2\pi R H u_R\right)}{\partial R}.
\end{equation}
In discredited form,
\begin{equation}
  \label{eq:ediscret} \frac{dq}{dt}=\dot e + \Delta \left(\pi R u_R
  \Sigma E \right)+ 2\pi R \Delta R P \dot H + P \Delta \left(2\pi R
  u_R H_\textrm{ad}\right),
\end{equation}
where $H_{\rm ad}$ is the advected value of $H$, and this term
corresponds to the advected volume contributing to the 'PdV' work.

Now, given the specific internal energy $E$, where 
\begin{equation}
E=\frac{3}{2}\frac{kT}{\mu m_p}+\frac{4\sigma}{c\rho}T^4,
\end{equation}
and given hydrostatic equilibrium in the form 
\begin{equation}
\rho\frac{kT}{\mu m_p}+\frac{4\sigma}{3c}T^4=G\frac{M}{4\rho R^3}\Sigma^2,
\end{equation}
$\rho$ and $T$ can now be be computed, for given values of $E$ and
$\Sigma$.  The variation of $H$ in terms of $E$ and $\Sigma$ is given
by using the definition of the surface density
(Equation~\ref{eq:sigma})
\begin{equation}
\frac{dH}{H}=\frac{d\Sigma}{\Sigma}-\frac{d\rho}{\rho}.
\end{equation}
Using both the hydrostatic equilibrium, and the definition of the
internal energy $E$, we can express the variation in $\rho$ in terms of
variations in the surface density $\Sigma$ and specific internal
energy $E$ as
\begin{equation}
\frac{d\rho}{\rho}=\frac{8-6\eta}{8+\beta_P-7\eta}\frac{d\Sigma}{\Sigma}-\frac{4-3\beta_P}{8+\beta_P-7\eta}\frac{dE}{E}\;.
\end{equation}
With the combination of the last two equations we can eliminate
$d\rho/\rho$ to obtain
\begin{eqnarray}
\frac{dH}{H}&=&\frac{\beta_P-\eta}{8+\beta_P-7\eta}\frac{d\Sigma}{\Sigma}+\frac{4-3\beta_P}{8+\beta_P-7\eta}\frac{dE}{E},\\
&=& \frac{4\beta_P-4-\eta}{8+\beta_P-7\eta}\frac{d\Sigma}{\Sigma}+\frac{4-3\beta_P}{8+\beta_P-7\eta}\frac{de}{e},\\
&=&
\frac{4\beta_P^2-11\beta_P+8}{\beta_P^2+13\beta_P-16}\frac{d\Sigma}{\Sigma}+\frac{-3\beta_P^2+10\beta_P-8}{\beta_P^2+13\beta_P-16}\frac{de}{e},
\label{eq:dh}\\
\end{eqnarray}
where we use 
\begin{equation}
\eta=\frac{E_\textrm{gas}}{E},
\end{equation}
and
\begin{equation}
\beta_P=\frac{P_\textrm{gas}}{P},
\end{equation}
as the fractional contribution of the gas component to the specific
internal energy and pressure, respectively. These two variables are
related by
\begin{equation}
  \label{eq:eta}
  \eta=\frac{\beta_P}{2-\beta_P}.
\end{equation}
The specific internal energy and pressure are related by
\begin{equation}
  E=\frac{P}{\rho}\left(3-\frac{3}{2}\beta_P\right)\;.
\end{equation}

Equation~\ref{eq:dh}, multiplied by $H$ and taking the time-derivative
leads to (using Equations~(\ref{eq:eta}) and (\ref{eq:sigma}))
\begin{equation}
\label{eq:doth}
\dot H
=\underbrace{\frac{4\beta_P^2-11\beta_P+8}{\beta_P^2+13\beta_P-16}}_{C_\Sigma}\frac{\dot
\Sigma}{2\rho}+\underbrace{\frac{-3\beta_P^2+10\beta_P-8}{\beta_P^2+13\beta_P-16}}_{C_e}\frac{\dot
e}{2\pi R \rho \Delta R E}
\end{equation}

The coefficients $C_e$ and $-C_\Sigma$ are equal to $\frac{1}{2}$ for
$\beta_P \to 0$ and $\beta_P \to 1$, while going through a minimum for
$\beta_P\approx 0.75$ at values $0.38$ and $0.35$, respectively. Then
in both  the limiting cases of $\beta_P \rightarrow 0$ and $\beta_P
\rightarrow 1$, we find  $H\propto
\sqrt{e/\Sigma} \approx \sqrt{\Sigma c_s^2/\Sigma}=c_s$,
recovering the hydrostatic equilibrium.

Using Equation~(\ref{eq:doth}) in the energy
equation~(\ref{eq:ediscret}), we can write
\begin{eqnarray*}
  \frac{dq}{dt}&=&\dot e + \Delta \left(\pi R u_R \Sigma E \right)+
  2\pi R \Delta R P \dot H + P \Delta \left(2\pi R u_R
  H_\textrm{ad}\right)\\ &=& \dot e\left(1+C_e\frac{P}{\rho
  E}\right)+\Delta \left(\pi R u_R \Sigma E \right)+ \pi R \Delta R
  \frac{P}{\rho} C_\Sigma \dot \Sigma + 2\pi R P \Delta \left(u_R
  H_\textrm{ad}\right)\\ &=& \dot e\left(1+C_e\frac{P}{\rho
  E}\right)+\Delta \left(\frac{1}{2}\dot M E_\textrm{ad}\right)+ \pi R
  \Delta R \frac{P}{\rho} C_\Sigma \dot \Sigma + \frac{1}{2}P \Delta
  \left(\dot M \frac{1}{\rho_\textrm{ad}}\right),
\end{eqnarray*}
where we applied $\dot M = 2\pi \Sigma R u_R$ in the last step.

\section{Local stability analysis} \label{sect:stabanalysis}

The thermal stability analysis \citep[]{1976MNRAS.177...65P}, now
including advection shows thermal instability for  
\begin{equation}
\left(\frac{\partial \log Q^+}{\partial \log T}\right)_P -
\left(\frac{\partial \log \left( Q^-+Q_\textrm{ad}\right)}{\partial
    \log T}\right)_P>0. 
\end{equation}
Thermal instability occurs, if the increase(decrease) of temperature
compared to the equilibrium state leads to a stronger
increase(decrease) of the heating compared to the cooling while
keeping hydrostatic equilibrium.

Using the equilibrium condition ($Q^+=Q^-+Q_\textrm{ad}$) and
\begin{equation}
\epsilon_\textrm{ad}=\frac{Q_\textrm{ad}}{Q_\textrm{ad}+Q^-},
\end{equation}
we get the condition
\begin{equation}
\left(\frac{\partial \log Q^+}{\partial \log T}\right)_P -
\epsilon \left(\frac{\partial \log Q_\textrm{ad}}{\partial \log
    T}\right)_P-\left(1-\epsilon\right)\left(\frac{\partial \log
    Q^-}{\partial \log T}\right)_P>0.
\end{equation}

Along similar lines, viscous instability
\citep[e.g.][]{1974ApJ...187L...1L,1981ARA+A..19..137P} only occurs as long as 
\begin{equation}
\left(\frac{\partial \log \dot M}{\partial \log \Sigma}\right)_{P,T}<0.
\end{equation}
The disc is viscously unstable, if an
increase(decrease) of accretion rate does lead to a lower/higher
surface density. 

For the calculation of the criterion we use the hydrostatic equilibrium
\begin{equation}
  \log P-\log \frac{GM}{4\rho R^3}\Sigma^2=f_1(\rho,T,\Sigma)=0,
\end{equation}
and the stationary energy equation
\begin{equation}
  \log Q^+-\log \left(Q^-+Q_\textrm{ad}\right)=f_2(\rho,T,\Sigma)=0\;.
\end{equation}

With the contribution of gas and radiation to the total pressure (see
eq.~\ref{eq:pressure}) we can write the total variation of the
hydrostatic equilibrium and energy equation in terms of $\rho$, $T$
and $\Sigma$, generalising the method of \citet{2005MNRAS.356....1M}
for a non selfgravitating, optically thick accretion disc. We neglect
changes in $\chi_\textrm{ad}$ since $\chi_\textrm{ad}$ is expected to
vary only very little. 

We take the variation of the hydrostatic equilibrium ($f_1$)
\begin{equation}
   A \frac{d\rho}{\rho}+B\frac{dT}{T}+C\frac{d\Sigma}{\Sigma}=0
\end{equation}
and the stationary energy equation ($f_2$) 
\begin{equation}
  D \frac{d\rho}{\rho}+E\frac{dT}{T}+F\frac{d\Sigma}{\Sigma}=0
\end{equation}
with respect to $\rho$, $T$ and
$\Sigma$. The coefficients $A\dots E$ are
\begin{eqnarray}
  A&=&\frac{\partial f_1}{\partial \log \rho}=1+\beta_P\\
  B&=&\frac{\partial f_1}{\partial \log T}=4-3\beta_P\\
  C&=&\frac{\partial f_1}{\partial \log \Sigma}=-2\\
  D&=&\frac{\partial f_2}{\partial \log \rho}=A_\textrm{R}\left(1-\epsilon_\textrm{ad}\right)+\left(\beta_P-1\right)\left(1-2\epsilon_\textrm{ad}\right)\\
  E&=&\frac{\partial f_2}{\partial \log T}=\left(B_\textrm{R}-4\right)\left(1-\epsilon_\textrm{ad}\right)+\left(4-3\beta_P\right)\left(1-2\epsilon_\textrm{ad}\right)\\
  F&=&\frac{\partial f_2}{\partial \log \Sigma}=2\left(1-\epsilon_\textrm{ad}\right)
\end{eqnarray}
with
\begin{equation}
\left(A_\textrm{R},B_\textrm{R}\right)=\frac{\partial \log \kappa_\textrm{R}}{\partial \log \left(\rho,T\right)}.
\end{equation}
Thermal instability exists, if 
\begin{equation}
AE-BD>0
\end{equation}
where we used the variation of the energy equation and expressed the
density variation $d\rho/\rho$ in terms of temperature $dT/T$ using the
variation of the hydrostatic equilibrium while keeping the surface
density $\Sigma$=const.
 
We have viscous instability for
\begin{equation}
-\left(\beta_P-1\right)\frac{CE-BF}{AE-BD} -\left(4-3\beta_P\right)\frac{AF-CD}{AE-BD} +1<0
\end{equation}
where we both expressed the density and temperature variations in
terms of the surface density variations and use this expression in the
angular momentum equation
\begin{equation}
\dot M - 3\pi \nu_\textrm{t} \Sigma=0
\end{equation}
to get the instability criterion.

In the radiation pressure ($\beta_P=0$) and Thomson scattering ($A_\textrm{R}=B_\textrm{R}=0$) dominated regime we get thermal instability only if 
\begin{equation}
4-12\epsilon_\textrm{ad}>0
\end{equation}
 and viscous instability as long as 
\begin{equation}
\frac{7\left(\epsilon_\textrm{ad}-1\right)}{1-3\epsilon_\textrm{ad}}<0
\end{equation}
Since the nominator of the LHS of the last expression is always
negative ($0<\epsilon_\textrm{ad}<1 $), the condition of thermal
instability and viscous instability are the same. 

To conclude, thermal and viscous instability only occurs, if
$\epsilon_\textrm{ad} < \frac{1}{3}$. A radiation pressure dominated
disc can be thermally and viscously unstable as long as advection only
contributes less than one third of the local energy loss. This clearly
shows the stabilising effect of energy advection on radiation pressure
dominated accretion discs. 

For bound-free and free-free absorption ($A_\textrm{R}=1$,
$B_\textrm{R}=-\frac{5}{2}$), it can be shown that in this case the
disc is stable with respect to thermal and viscous instability for all
$\beta_P$.

It needs to be stated that these results are only based on a local
stability analysis and it still needs to be shown by either
time-dependent simulations or a global stability analysis when these
unstable modes are operational.

\label{lastpage}

\end{document}